\documentstyle[psfig]{mn}

\newif\ifAMStwofonts

\newcommand{\spose}[1]{\hbox to 0pt{#1\hss}}
\newcommand{\approxpropto}{\mathrel{\spose{\lower 3pt\hbox{$\sim$}}
	\raise 2.0pt\hbox{$\propto$}}}
\def\approxgt{\mathrel{\spose{\lower 3pt\hbox{$\sim$}}
	\raise 2.0pt\hbox{$>$}}}
\def\approxlt{\mathrel{\spose{\lower 3pt\hbox{$\sim$}}
	\raise 2.0pt\hbox{$<$}}}

\def \kmsMpc{\,\hbox{kms}^{-1}\,\hbox{Mpc}^{-1}}
\def \hkpc{\,h^{-1} \hbox{kpc}}
\def \hMpc{\,h^{-1} \hbox{Mpc}}
\def \hMsol{\,h^{-1} \hbox{M}_{\odot}}

\def \h2gcm3{h^{2} \, \hbox{g}\,\hbox{cm}^{-3}}
\def \Lxunit{\, \hbox{h}^{-2} \, \hbox{erg} \, \hbox{s}^{-1}}
\def \kevcm2{\, \hbox{keV} \, \hbox{cm}^{2}}
\def \cm3{\, \hbox{cm}^{-3}}
\def \Msunperyear{\,\hbox{M}_{\odot}\,\hbox{yr}^{-1}}

\ifoldfss
  \ifCUPmtlplainloaded \else
    \NewTextAlphabet{textbfit} {cmbxti10} {}
    \NewTextAlphabet{textbfss} {cmssbx10} {}
    \NewMathAlphabet{mathbfit} {cmbxti10} {} 
    \NewMathAlphabet{mathbfss} {cmssbx10} {} 
  \fi
  \ifAMStwofonts
    \ifCUPmtlplainloaded \else
      \NewSymbolFont{upmath} {eurm10}
      \NewSymbolFont{AMSa} {msam10}
      \NewMathSymbol{\upi}     {0}{upmath}{19}
      \NewMathSymbol{\umu}     {0}{upmath}{16}
      \NewMathSymbol{\upartial}{0}{upmath}{40}
      \NewMathSymbol{\leqslant}{3}{AMSa}{36}
      \NewMathSymbol{\geqslant}{3}{AMSa}{3E}

    \fi
  \fi
\fi 

\ifnfssone
  \newmathalphabet{\mathit}
  \addtoversion{normal}{\mathit}{cmr}{m}{it}
  \addtoversion{bold}{\mathit}{cmr}{bx}{it}
  \newmathalphabet{\mathbfit} 
  \addtoversion{normal}{\mathbfit}{cmr}{bx}{it}
  \addtoversion{bold}{\mathbfit}{cmr}{bx}{it}
  \newmathalphabet{\mathbfss} 
  \addtoversion{normal}{\mathbfss}{cmss}{bx}{n}
  \addtoversion{bold}{\mathbfss}{cmss}{bx}{n}
  \ifAMStwofonts
    \ifCUPmtlplainloaded \else
      \UseAMStwoboldmath
      \makeatletter
      \new@mathgroup\upmath@group
      \define@mathgroup\mv@normal\upmath@group{eur}{m}{n}
      \define@mathgroup\mv@bold\upmath@group{eur}{b}{n}
      \edef\UPM{\hexnumber\upmath@group}
      \new@mathgroup\amsa@group
      \define@mathgroup\mv@normal\amsa@group{msa}{m}{n}
      \define@mathgroup\mv@bold\amsa@group{msa}{m}{n}
      \edef\AMSa{\hexnumber\amsa@group}
      \makeatother
      \mathchardef\upi="0\UPM19
      \mathchardef\umu="0\UPM16
      \mathchardef\upartial="0\UPM40
      \mathchardef\leqslant="3\AMSa36
      \mathchardef\geqslant="3\AMSa3E
    \fi
  \fi
\fi 

\ifnfsstwo
  \DeclareMathAlphabet{\mathbfit}{OT1}{cmr}{bx}{it}
  \SetMathAlphabet\mathbfit{bold}{OT1}{cmr}{bx}{it}
  \DeclareMathAlphabet{\mathbfss}{OT1}{cmss}{bx}{n}
  \SetMathAlphabet\mathbfss{bold}{OT1}{cmss}{bx}{n}
  \ifAMStwofonts
    \ifCUPmtlplainloaded \else
      \DeclareSymbolFont{UPM}{U}{eur}{m}{n}
      \SetSymbolFont{UPM}{bold}{U}{eur}{b}{n}
      \DeclareSymbolFont{AMSa}{U}{msa}{m}{n}
      \DeclareMathSymbol{\upi}{0}{UPM}{"19}
      \DeclareMathSymbol{\umu}{0}{UPM}{"16}
      \DeclareMathSymbol{\upartial}{0}{UPM}{"40}
      \DeclareMathSymbol{\leqslant}{3}{AMSa}{"36}
      \DeclareMathSymbol{\geqslant}{3}{AMSa}{"3E}
    \fi
  \fi
\fi 

\ifCUPmtlplainloaded \else
  \ifAMStwofonts \else 
    \def\upi{\pi}
    \def\umu{\mu}
    \def\upartial{\partial}
  \fi
\fi

\title[Simulations of the ICM]
{Cosmological simulations of the intracluster medium}
\author[S.T. Kay et al.]
       {Scott T. Kay,$^{1}$\thanks{E-mail: s.t.kay@sussex.ac.uk}
	Peter A. Thomas,$^{1}$ Adrian Jenkins$^{2}$ and Frazer R. Pearce$^{3}$\\ 
        $^{1}$ Department of Physics and Astronomy, University of Sussex, Falmer, 
	Brighton BN1 9QH\\	
	$^{2}$ Institute for Computational Cosmology, Physics Department, 
        University of Durham, South Road, Durham DH1 3LE\\
	$^{3}$ Physics and Astronomy Department, University of Nottingham, Nottingham NG7 2RD\\
	}
\date{Accepted .
      Received ;
      in original form }

\pagerange{\pageref{firstpage}--\pageref{lastpage}}
\pubyear{2002}

\begin{document}

\maketitle

\label{firstpage}

\begin{abstract}
We investigate the properties of the intracluster medium (ICM) that forms within 
$N$-body/hydrodynamical simulations of galaxy clusters in a $\Lambda$CDM cosmology.
When radiative cooling and a simple model for galactic feedback
are included, our clusters have X-ray luminosities and temperatures 
in good agreement with observed systems, demonstrating the required 
excess entropy in their cores. More generally, cooling and feedback 
increases the entropy of the ICM everywhere, albeit without 
significantly affecting the slope of the profile ($S \propto r$) at large radii. 
The temperature of the ICM is only modestly increased by these
processes, with projected temperature profiles being in reasonable
agreement with the observations. 
Star/galaxy formation is still too efficient in our simulations, however, 
and so our gas mass fractions are around 60 per cent of the observed
value at $r_{2500}$. 
Finally, we examine the reliability of using the hydrostatic equilibrium equation
to estimate cluster masses and find that it underpredicts the true mass
of our clusters by up to 20 per cent, due to incomplete thermalisation of the gas.
Feedback reduces this discrepancy, however, with estimates being accurate to within 
10 per cent out to $r_{500}$.
\end{abstract}


\section{Introduction}

Clusters of galaxies play an important role in our quest for
determining the cosmology of our Universe. 
In models of hierarchical structure formation such as the $\Lambda$CDM
model, the mass function of clusters is a sensitive
function of cosmological parameters, notably the matter density parameter,
$\Omega_{\rm m}$ and the amplitude of linear fluctuations on 8$\hMpc$ scales, $\sigma_8$. 
As a result, many authors have combined the observed (in the X-ray) 
abundance of clusters with
theoretical mass functions (Press \& Schechter 1974; Sheth, Mo \& Tormen 2001;
Jenkins et al. 2001) to constrain these parameters
(e.g. Evrard 1989; Henry \& Arnaud 1991; Oukbir \& Blanchard 1992; 
White, Efstathiou \& Frenk 1993; 
Viana \& Liddle 1996; Eke, Cole \& Frenk 1996; Henry 2000; Pierpaoli et al. 2003; 
Viana et al. 2003).

In the post-{\it WMAP} era (Spergel et al. 2003; Verde 2003), particular
attention is now being paid to systematic uncertainties that may bias these estimates.
As was recently pointed out by Henry (2004), the largest
systematic uncertainty in determining $\sigma_8$ using X-ray clusters is in the 
calibration of the X-ray temperature to mass relation.
In order for robust determinations of such quantities to
be made, accurate theoretical predictions for how clusters form and evolve are
required, taking full account of the physics that governs the structure of
the intracluster medium (ICM).

The best tool we have at our disposal to model the ICM is direct 
numerical simulation. Evrard (1990) performed the first cosmological simulation of a 
cluster and in his pioneering study, demonstrated that many of its
properties resembled those derived from X-ray observations. Subsequent 
studies that followed (e.g. Thomas \& Couchman 1992; Katz \& White 1992; Cen \& Ostriker 1994;
Navarro, Frenk \& White 1995; Evrard, Metzler \& Navarro 1996; 
Bryan \& Norman 1998; Eke, Navarro \& Frenk 1998;
Frenk et al. 1999; Thomas et al. 2001) have significantly improved our understanding of X-ray 
clusters within hierarchical models.

Many of the first cluster simulations ignored radiative cooling of the gas
on the grounds that clusters today have average cooling times longer than the age
of the universe. These `non-radiative' simulations
vindicated simple self-similar scaling relations, as expected when the structure
of the ICM is determined solely by gravity (Kaiser 1986). 
As useful as non-radiative simulations have been however, they contradict X-ray observations
of clusters. The clearest example is the luminosity-temperature ($L_{\rm X}-T_{\rm X}$)
relation, that is observed to have a steeper slope ($L_{\rm X} \propto T_{\rm X}^{\sim 3}$, 
e.g. Edge \& Stewart 1991) than predicted from gravitational 
heating models ($L_{\rm X} \propto T_{\rm X}^2$). Another useful viewpoint 
is that the ICM is less concentrated
because its entropy is higher (Kaiser 1991; Evrard \& Henry 1991),
as demonstrated by Ponman, Cannon \& Navarro (1999). Consequently, much
attention has been paid to models which focus on the entropy of the ICM
(see e.g. Bower 1997; Voit \& Bryan 2001; Voit et al. 2002,2003a). 

Early models concentrated on the {\it preheating} hypothesis, that the ICM was heated
externally when the density contrast was low, therefore requiring less energy to
reach the desired entropy level than if the heating occurred in a denser environment.
Simulations of preheating models have
been shown to be capable of producing the desired effect on the overall entropy level of the ICM 
(e.g. Navarro et al. 1995; Bialek, Evrard \& Mohr 2001; Borgani et al. 2002)
but have since fallen out of favour for many reasons, for example they predict
isentropic cores in low-mass clusters that are not observed 
(see Ponman, Sanderson \& Finoguenov 2003).

More recently, attention has shifted to modelling the effects of radiative cooling on the ICM. 
Naively, since the entropy of cooling gas decreases, one would expect cooling to
have the opposite of the desired effect on the ICM, making clusters even more luminous
than when cooling is neglected. Indeed, this effect was seen in some numerical studies
(Katz \& White 1992; Suginohara \& Ostriker 1998; Lewis et al. 2000), but was caused by the 
presence of an unrealistically-large central galaxy. Pearce et al. (2000) demonstrated that 
this `overcooling' problem could be ameliorated by 
`decoupling' the hot gas from the cold galactic gas, and as a result, found that
cooling actually increased the ICM entropy. Most of the galactic material
cools very quickly at high redshift without ever reaching the virial temperature
(Binney 1977; Kay et al. 2000; Binney 2004), and instead emits in the Ultra-Violet (Fardal et al. 2000).
The ICM adjusts itself to the loss of pressure support, with higher entropy gas flowing
inwards to replace the cooled gas (Pearce et al. 2000). Various subsequent studies of cooling
have successfully reproduced various observed X-ray properties of clusters (e.g. Bryan 2000; 
Muanwong et al. 2001,2002; Wu \& Xue 2002; Dav\'{e}, Katz \& Weinberg 2002; Thomas et al. 2002; Voit et al.
2002).

Although cooling contributes to the excess entropy of the ICM, heating processes (from
stars and/or active galactic nuclei) are
still required in order to regulate galaxy formation and possibly shape their
luminosity function (e.g. White \& Frenk 1991; Cole 1991; Balogh et al. 2001; 
Benson et al. 2003; Binney 2004). Numerical efforts are now focusing on incorporating such
feedback processes with cooling into simulations of groups and clusters 
(Kay, Thomas \& Theuns 2003; Tornatore et al. 2003; Kay 2004; Borgani et al. 2004)
and will continue to remain a vital area of study for the foreseeable future, until
the effects of galaxy formation can be successfully incorporated into ICM models.

In this paper we focus on a set of high-resolution cosmological 
simulations of 15 galaxy clusters,
with the primary aim of investigating how the combined effects of radiative
cooling and a simple model for targeted heating by galactic feedback affects 
the structure of the ICM and as a consequence, estimates of the 
total mass distribution in clusters. 
Our study is in many ways similar to, and can be viewed as a progression of,
those performed by Ascasibar et al. (2003) and Rasia, Tormen \& Moscardini (2004), 
who restricted their simulations to non-radiative systems. 
We compare our findings with their results and with observational data
where appropriate.

The rest of this paper is organised as follows. In Section~\ref{sec:sims}
we outline our method and discuss general properties of our simulated clusters. 
Radial profiles
derived from the distribution of cluster mass (total mass density, ICM density,
entropy and gas fraction) are discussed in Section~\ref{sec:massprof}. 
In Section~\ref{sec:temprof} we present cluster temperature profiles. Cluster 
mass determinations are investigated under various conditions in Section~\ref{sec:mest}. 
Finally, we draw conclusions in Section~\ref{sec:conclusions}.

\section{Simulation details}
\label{sec:sims}

Simulations were performed assuming a $\Lambda$CDM cosmology, setting
$\Omega_{\rm m}=0.3, \Omega_{\Lambda}=0.7, \Omega_{\rm b}=0.045, h=0.7$ 
\& $\sigma_8=0.9$. Our choice of parameters are in reasonably good agreement
with the {\it WMAP} results (Spergel et al. 2003). 

All simulations were performed with version 2 of the {\sc gadget} $N$-body/hydrodynamics
code (Springel, Yoshida \& White 2001), kindly made available to us by V.~Springel.
Gravitational forces were computed using the combination of particle-mesh and tree algorithms.
Hydrodynamical forces were calculated using the variant of the Smoothed Particle Hydrodynamics
(SPH) algorithm proposed by Springel \& Hernquist (2002) that explicitly conserves entropy
where the (artificial) viscosity of the flow is zero.

Two sets of simulations were used in this study. First of all, we performed
simulations of a random volume, sampling a cube of comoving length,
$L=120 \hMpc$ using $N=2\times 256^3$ particles (half baryons, half dark matter). 
The dark matter and baryon particle masses are then  $m_{\rm dark} = 7.3 \times 10^{9} \hMsol$
and $m_{\rm bary} = 1.3 \times 10^{9} \hMsol$ respectively. 
The second set of simulations consisted of 10 resimulated clusters, 
selected from a large $N$-body simulation performed by the Virgo
Consortium (Yoshida, Sheth \& Diaferio 2001), with $N=512^3$ and $L=479\hMpc$. 
The ten most massive objects with virial masses below $10^{15}\hMsol$ were selected
and resimulated at higher resolution, with similar particle masses to the random
volume simulations.
This sample has already been studied elsewhere without gas (see e.g. Gao et al. 2004). 

For all simulations, we started the runs at $z=49$ and evolved them to $z=0$ (our
results are presented for this redshift). We fixed
the gravitational softening in comoving co-ordinates at all times to an
equivalent Plummer value of $\epsilon = 20 \hkpc$ ({\sc gadget} uses a spline
softening kernel where the force becomes exactly Newtonian for $r>2.8\epsilon$).

\subsection{The models}

In this paper we focus on two models, differing only in the freedom
allowed for the entropy 
\footnote{We define entropy as $S=kT\,(\rho/\mu m_{\rm H})^{1-\gamma}$,
where $\gamma=5/3$ is the ratio of specific heats for a monatomic
ideal gas and $\mu m_{\rm H}=0.6$ is the mean atomic weight of
a fully-ionised plasma.}
of the gas to vary. 
In the first model, hereafter
referred to as the {\it Non-Radiative} model, 
entropy can only increase due to shock-heating.
Although it has already been demonstrated
that this model does not produce clusters that resemble observed systems 
(e.g. Evrard \& Henry 1991) it nevertheless has been a well studied model
and provides a good reference to measure the effects of non-gravitational
processes on the ICM.

In the second model, hereafter the {\it Feedback}
model, we included two additional physical processes, both thought
to be crucial in producing realistic clusters. First of all, the gas
was allowed to lose entropy through radiative cooling. Although clusters
at low redshift have average cooling times that are longer than the age of 
the Universe, the same cannot be said for their precursors at high redshift 
when most of their galaxies form. The result of this removal of low entropy gas
is a net increase in the entropy of the remaining ICM (Pearce et al. 2000). 
We included radiative cooling using the same method
as used in the {\sc hydra} code (see Thomas \& Couchman 1992), adopting
a tabulated cooling function (Sutherland \& Dopita 1993) with $Z=0.3Z_{\odot}$.

Gas is able to cool down to $10^4$K, where we arbitrarily cut off the cooling
rate. The physics that governs the fate of this cooled gas, now
identified as galactic material (e.g. Kay et al. 2000), is complex and would
not be resolved in our simulations (although see e.g. Springel \& Hernquist 2003 
for an attempt to model a multiphase galactic component within cosmological 
simulations). We instead adopt a phenomenological approach to galaxy formation,
and assume that a fixed mass fraction of cooled gas will go on to form stars
and the rest will be reheated by the stars
(the heating could also be due to active galactic nuclei, 
but note our model does not allow us to discriminate between the two).

First, we identify cooled gas using a density threshold, $n_{\rm H}>n_{*}$
(where $n_{\rm H}=X \rho /m_{\rm H}$ is the hydrogen density
and $X=0.76$ the hydrogen mass fraction)
and a temperature threshold, $T<T_{*}$, setting $n_{*}=10^{-3} \cm3$ and 
$T_{*}=1.2\times 10^{4}$K. Although typical densities
where star/black hole formation occurs are much higher, we deliberately set the threshold to be
low as it ensures that a negligible amount of thermal energy in the reheated gas is
lost as radiation. Note therefore that we will be ignoring some
cold gas that would otherwise be ablated by the ICM.

Whether a particle then forms stars or is reheated is determined on a stochastic basis.
The mass fraction of reheated gas is controlled by a parameter $f_{\rm heat}$ (hence
a fraction, $1-f_{\rm heat}$ will form stars).
For each cooled gas particle we draw a random number, $r$, from the unit interval
and increase its entropy by a fixed amount, $S_{\rm heat}$, if $r<f_{\rm heat}$. 
The temperature of the reheated gas is then 
\begin{equation}
kT_{\rm heat} =     \left( {S_{\rm heat} \over 100 \kevcm2}  \right) \,
	            \left( {n_{\rm heat} \over 10^{-3} \cm3} \right)^{2/3} \, 
                    \hbox{keV}.
\label{eqn:theat}
\end{equation}
We adopted a choice of parameters equivalent to that used by Kay (2004)
in his {\it Strong Feedback} model, namely $S_{\rm heat}=1000\kevcm2$
and $f_{\rm heat}=0.1$. For gas with density, $n_{\rm H}=n_{*}$,
$kT_{\rm heat} = 17$ keV, although part of this energy is distributed
as the gas does work on its neighbours. 

This energy is only carried by 10 per cent of the mass of cooled gas, so it is
useful to estimate the overall energy deposited into the ICM. 
Such an estimate is
\begin{equation}
E_{\rm ICM} = {3 \over 2} \, {kT_{\rm heat} \over \mu m_{\rm H}} \,
M_{\rm heat},
\label{eqn:eheat1}
\end{equation}
where $M_{\rm heat}=(f_{\rm heat}^{-1}-1)^{-1}M_{*}$ is the mass
of reheated gas, directly related to the residual mass of stars
in the cluster, $M_{*}$. For our choice of parameters, 
$E_{\rm ICM} \sim 10^{63} (M_{*}/10^{14}\hbox{M}_{\odot})$ erg,
comparable with the maximum energy available from Type~II supernovae, 
assuming one supernova releases $10^{51}$ erg and the number of Type~II supernovae,
$N_{\rm SNII}=0.01(M_{*}/M_{\odot})$. Thus, on average, the energy added to each 
ICM atomic nucleus is $kT_{\rm ICM} = 3.1 (f_{*}/f_{\rm gas})$ keV, where
$f_{*}$ and $f_{\rm gas}$ are the star and ICM mass fractions
respectively. For clusters in our {\it Feedback} model, we find 
$kT_{\rm ICM} \sim 1$keV, i.e. a similar level to that found in previous
studies including heating 
(e.g. Wu, Fabian \& Nulsen 2000; Bower et al. 2001; Borgani et al. 2001; 
Muanwong et al. 2002).

\subsection{Cluster identification}

Clusters were identified using the procedure detailed in Muanwong et al. (2002).
This involved finding the cluster centre (by searching for the density maximum)
then growing spheres around this centre until the average density reached
a constant factor, $\Delta$, times the critical density. This threshold defines the 
relation between cluster mass and radius
\begin{equation}
M_{\Delta} = {4\pi \over 3} \, \Delta \, \rho_{\rm cr} \, R_{\Delta}^3,
\label{eqn:massdef}
\end{equation}
where $\rho_{\rm cr}=3H_0^2/8\pi G$ is the critical density and
$H_0=70 \kmsMpc$ for our cosmology.
We define the virial mass/radius using $\Delta=178\,\Omega_{\rm m}^{0.45} \sim 104$
(Eke et al. 1998). Our main
results will refer to virial quantities but occasionally
we will use other values of $\Delta$ where appropriate.

We only considered clusters with $M_{\rm vir}>4.3\times 10^{14}\hMsol$, 
corresponding to an effective particle number threshold,
$N = M_{\rm vir}/(m_{\rm dark}+m_{\rm gas}) > 50,000$, as suggested
by Borgani et al. (2002) for obtaining reliable X-ray properties.
All our resimulated
clusters and the 5 most massive objects in our random volume simulations
satisfied this criterion, bringing our sample to 15 clusters per model.

\subsection{Cluster sample and global properties}

\begin{table}
\caption{Global parameters of the simulated cluster sample,
from left to right: 
cluster number; 
virial radius, in $\hMpc$; 
virial mass in $10^{14}\hMsol$; 
virial temperature in keV;
number of baryon particles (gas or stars) within $R_{\rm vir}$;
ICM gas mass fraction; stellar mass fraction}
\begin{center}
\begin{tabular}{rrrrrrr}
\hline
No. & $R_{vir}$ & $M_{vir}$ & $kT_{vir}$ & $N_{\rm bary}$ & $f_{\rm gas}$ & $f_{*}$\\
\hline
1  & 2.41 & 16.83 & 7.70 & 184,082 & 0.105 & 0.035\\
2  & 2.03 & 10.08 & 4.91 & 92,833  & 0.102 & 0.032\\
3  & 2.02 &  9.82 & 6.06 & 88,840  & 0.097 & 0.035\\
4  & 1.99 &  9.44 & 5.57 & 86,347  & 0.098 & 0.035\\
5  & 1.99 &  9.43 & 6.00 & 89,872  & 0.105 & 0.034\\
6  & 1.96 &  9.03 & 5.10 & 84,546  & 0.103 & 0.033\\
7  & 1.94 &  8.68 & 6.57 & 82,922  & 0.104 & 0.035\\
8  & 1.82 &  7.24 & 4.34 & 75,390  & 0.098 & 0.036\\
9  & 1.80 &  6.93 & 4.73 & 68,042  & 0.108 & 0.035\\
10 & 1.79 &  6.91 & 4.66 & 62,766  & 0.097 & 0.035\\
11 & 1.78 &  6.80 & 4.97 & 64,532  & 0.104 & 0.034\\
12 & 1.73 &  6.22 & 4.02 & 56,027  & 0.096 & 0.035\\
13 & 1.70 &  5.88 & 3.74 & 62,316  & 0.100 & 0.036\\
14 & 1.67 &  5.56 & 4.54 & 56,040  & 0.091 & 0.038\\
15 & 1.59 &  4.77 & 3.99 & 48,156  & 0.091 & 0.038\\
\hline
\end{tabular}
\label{tab:sample}
\end{center}
\end{table}

Table~\ref{tab:sample} lists global properties of our 15 
{\it Feedback} clusters. The systems span a factor of 4 in mass,
with $kT_{\rm vir} \sim 4-8$keV.
We define the ICM as all gas particles within $R_{\rm vir}$ with
$kT > 0.1$keV (i.e. $\sim 1.2 \times 10^{6}$K). Generally, very little
or no gas exists in our clusters below this temperature threshold
(in the {\it Feedback} model, a few gas particles exist that have 
cooled down to $T=10^4$K but have not yet formed stars or been reheated).

There is very little scatter between ICM gas fraction values,  
$f_{\rm gas}=0.100 \pm 0.005$ (for the $Non-Radiative$ clusters, 
$f_{\rm gas}=0.136 \pm 0.004$). We note the average
star fraction, $f_{*} = 0.035 \pm 0.002$, i.e. 
$\sim 25$ per cent of the baryon mass is collisionless. By construction,
this fraction is lower than would be expected in the absence of feedback,
however it is still higher than the observed stellar mass
fraction, measured to be $\approxlt 15$ per cent (e.g. Balogh et al. 2001;
Lin, Mohr \& Stanford 2003). The observed value is an
underestimate of $f_{*}$ as part of this mass may be hidden in a diffuse
component (e.g. Murante et al. 2004), however note that the desired amount of 
cooled baryons can be achieved in our model by increasing $f_{\rm heat}$,
without significantly affecting the global properties of the ICM
(Kay et al. 2003).

\begin{figure}
\centering
\centerline{\psfig{file=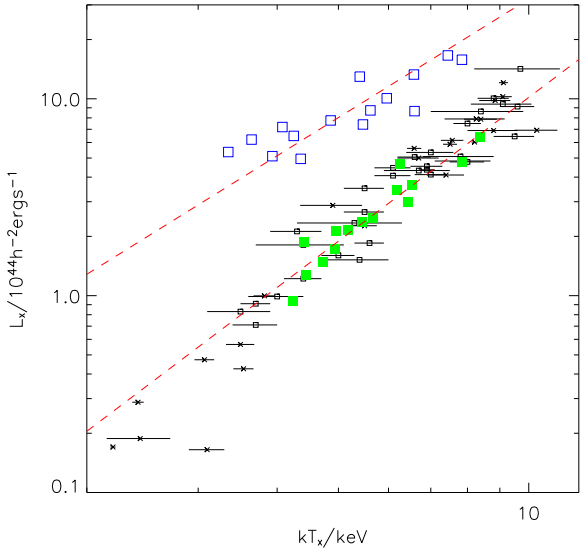,width=8.5cm}}
\caption{X-ray luminosity-temperature relation for clusters in the
{\it Non-Radiative} (open squares) and {\it Feedback} (filled squares) 
models. Dashed lines are power law fits to the data (assuming 
$L_{\rm X} \propto T_{\rm X}^2$ for the {\it Non-Radiative} clusters). 
Symbols with error bars are for clusters studied by Markevitch (1998b) and
Arnaud \& Evrard (1998).}
\label{fig:lxtx}
\end{figure}

To illustrate the difference between the X-ray properties of
the clusters in the two models, we show in Fig.~\ref{fig:lxtx} 
the $L_{\rm X}-T_{\rm X}$ relation 
at $z=0$. {\it Non-Radiative} clusters are represented by
open squares and {\it Feedback} clusters filled squares. 
X-ray emission-weighted temperatures and luminosities were computed 
using the procedure outlined by Muanwong et al. (2002). 
To be approximately consistent with the observational data, 
luminosities are bolometric and temperatures calculated in a
hard (1.5-8.0 keV) X-ray band. Furthermore, emission from within 
50$\hkpc$ of the cluster centre was omitted 
(i.e. we removed excess emission present due to cooling flows).

The upper dashed line illustrates a best-fit power law to the 
{\it Non-Radiative} clusters, assuming $L_{\rm X} \propto T_{\rm X}^2$.
This is the expected scaling relation for self-similar clusters assuming
the dominant emission mechanism is thermal bremsstrahlung (Kaiser 1986); our
clusters are in reasonable agreement with this scaling.

Cooling and feedback processes combine to reduce the emission from within each cluster, 
particularly in low-temperature systems (see Kay 2004). 
This lowers and steepens the $L_{\rm X}-T_{\rm X}$ 
relation, bringing the {\it Feedback} clusters into good agreement with the 
observational data 
(Markevitch 1998b; Arnaud \& Evrard 1998). The best-fit relation for these systems
is $\log(L_{\rm X,44}) = (-1.42\pm 0.17) + (2.43 \pm 0.23)\log(kT_{\rm X}/\hbox{keV})$, 
where $L_{\rm X,44}$ is $L_{\rm X}$ in units of $10^{44} \, \Lxunit$.

Recently, it has emerged from X-ray spectra that there is a lack of cool gas 
(below one third of the mean temperature) in the cores of clusters, even though 
the temperature of the gas is decreasing radially inwards and its cooling time is 
significantly shorter than a Hubble time (e.g. Fabian 2003). Such spectroscopic 
mass-deposition rates, $\dot{M}_{\rm spec}$, are of order 10
$\Msunperyear$, about a factor of 5-10 lower than rates estimated directly 
from the core X-ray emission of a cluster
\begin{equation}
\dot{M}_{\rm X} = {2 \over 5} \, {L_{\rm X} \, \mu m_{\rm H} \over kT_{\rm X}}.
\label{eqn:mdot}
\end{equation}
Unfortunately our model clusters do not have the resolution for us
to study their cooling flows in detail (a $10 \Msunperyear$ cooling flow corresponds
to one gas particle cooling approximately every 200 Myr), in particular, to
measure $\dot{M}_{\rm spec}$. However, we can estimate the apparent mass deposition
rate, $\dot{M}_{\rm X}$.
 \begin{figure}
\centering
\centerline{\psfig{file=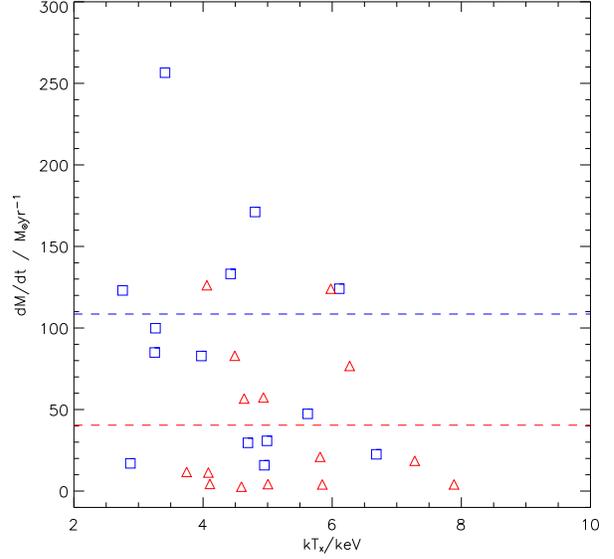,width=8.5cm}}
\caption{Mass deposition rates for the {\it Non-Radiative} (squares)
and {\it Feedback} (triangles) clusters, inferred from their core X-ray
emission, as a function of emission-weighted temperature. Dashed lines
give the mean rates.}
\label{fig:mdot}
\end{figure}
Fig.~\ref{fig:mdot} illustrates such rates for our {\it Non-Radiative}
and {\it Feedback} clusters (calculated within $50\hkpc$ of the cluster centre).
Note that both models exhibit a large spread in values:
in the former case, $\dot{M}_{\rm X}$ ranges from $20-400 \Msunperyear$
(although the gas can never cool),
while in in the latter case, $\dot{M}_{\rm X} = 3-130 \Msunperyear$.
The {\it Feedback} model, with $\left<\dot{M}_{\rm X}\right> = 40\Msunperyear$,
is in broad agreement with the observations.


\section{Cluster mass profiles}
\label{sec:massprof}

Measuring the mass (both baryonic and total) distribution
in clusters is of primary interest as it allows constraints to
be placed on cosmological parameters, for example by 
calibrating the relation between mass and X-ray temperature
(e.g. Henry 2004) or luminosity (e.g. Allen et al. 2003), or 
by using the gas mass fraction itself
(e.g. White et al. 1993b; Evrard 1997;
Allen et al. 2002,2004). Knowledge of the ICM temperature distribution 
allows gas mass density profiles to be estimated via deconvolution of X-ray 
surface brightness profiles, and total masses estimated via the
equation of hydrostatic equilibrium. We will return
to the latter issue in Section~\ref{sec:mest}, but for now, 
examine the mass distribution of our simulated clusters directly.

\subsection{Mass density profiles}

\begin{figure}
\centering
\centerline{\psfig{file=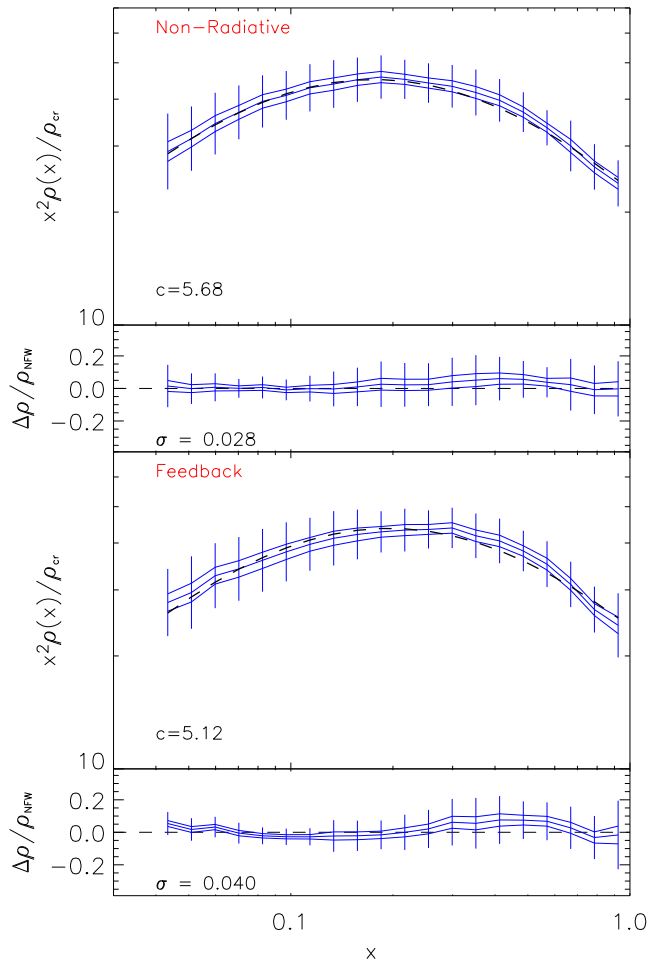,width=8.5cm}}
\caption{Total mass density profiles, $x^2\rho(x)$, where
$x=r/r_{\rm vir}$, for clusters in the 
{\it Non-Radiative} and {\it Feedback} models. Solid curves are 
the mean profile and the error on the mean, and the error bars illustrate 
the standard deviation within each
bin. The dashed curve is the best-fit NFW profile to the mean curve,
weighted by the error on the mean, with the
concentration parameter given in the legend. Also
plotted below each main panel is the fractional differences between
individual profiles and corresponding best-fit NFW profiles. Again, the solid
curve is the mean and the error bars the standard deviations within each bin.
The rms fractional difference, averaged over all bins, is given in the legend.}
\label{fig:denproftot}
\end{figure}

Shown in Fig.~\ref{fig:denproftot} are total mass density profiles, 
plotted as $x^2\rho(x)/\rho_{\rm cr}$, where $\rho_{\rm cr}$ is the critical
density and $x=r/r_{\rm vir}$. We set the minimum radius to be $x_{\rm min}=0.04$ 
as this radius is both larger than the spline gravitational softening length 
and contains at least 100 hot gas particles for all clusters in our sample
(Borgani et al. 2002; Ascasibar et al. 2003).
Solid curves represent the mean profile and the error on the mean while error
bars illustrate standard deviations within each bin, highlighting cluster
to cluster variations.
The scatter in density about the mean
at each radius is quite small for both models, $ \sim \pm 15$ per cent when 
averaged over all bins. 

A useful one-parameter model was proposed by Navarro et al.
(1995, 1997, hereafter NFW)
\begin{equation}
{\rho(x) \over \rho_{\rm cr}} = {\delta_c \over cx (1 + cx)^2},
\label{eqn:rho_nfw1}
\end{equation}
where $c$ is the concentration parameter and
\begin{equation}
\delta_c = {\Delta \over 3} \, {c^3 \over \ln (1+c)-c/(1+c)},
\label{eqn:rho_nfw2}
\end{equation}
where $\Delta = 104$ in our case. Best-fit NFW profiles to the mean
density profiles, weighted by the error on the mean, 
are plotted as dashed curves in the figure. The mean
profile concentration is $5.7$ for the {\it Non-Radiative} model 
and $5.1$ for the {\it Feedback} model. Cooling and feedback
slightly flatten the inner profile ($x<0.3$), causing a $\sim 10$
per cent increase in the best-fit value of $c$.

In the lower panels of Fig.~\ref{fig:denproftot}, we plot the
fractional difference, $\Delta \rho/\rho$, between individual profiles and 
their corresponding NFW model fits. Again, the solid curve is the result
for the mean profile and the error bars the standard deviation within
each bin. Thomas et al. (2001) pointed out that, for their non-radiative
clusters simulated in a $\tau$CDM cosmology, the NFW model generally 
does not provide a good fit to individual systems but does well when 
fitting a profile averaged over clusters with similar mass. For our
clusters, the fit also improves significantly when averaging over
clusters. The mean deviation at a given radius is around 10 per cent,
but the root-mean-square difference, 
$\sqrt{\left< (\Delta \rho / \rho)^2 \right>}$, when averaged
over all clusters and bins is only a few per cent.

\subsection{ICM density profiles}
\label{subsec:denprof}

\begin{figure}
\centering
\centerline{\psfig{file=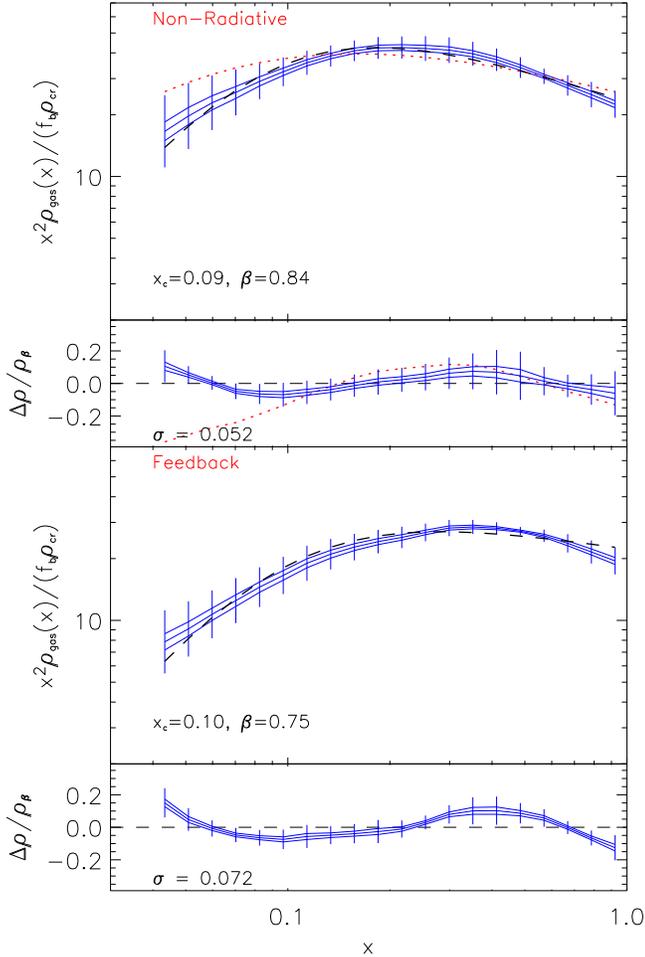,width=8.5cm}}
\caption{ICM density profiles, $x^2\rho_{\rm gas}(x)$, fitted
using the isothermal $\beta$-model.
The model profile fitted to the results
of Rasia et al. (2004) is shown as a dotted curve.}
\label{fig:denprofgas}
\end{figure}

Fig.~\ref{fig:denprofgas} illustrates ICM density profiles:
comparing the two simulation models reveals that the {\it Feedback} clusters have a 
lower density than the {\it Non-Radiative} clusters everywhere, with the difference 
being particularly pronounced in the central region. This is caused both by
cooling (which removes gas from the ICM) and feedback (which heats the ICM). 
Since the cooling rate scales as $\rho$ and the feedback rate effectively scales 
with the cooling rate, we expect their influence to be more pronounced closer to the 
cluster centre.

Historically, the isothermal $\beta$-model 
(Cavaliere \& Fusco--Femiano 1976) is used to fit X-ray
surface brightness profiles (we will return to this in Section~\ref{sec:mest}), 
with the underlying ICM density profile being
\begin{equation}
{\rho_{\rm gas}(x) \over f_{\rm b}\rho_{\rm cr}} = 
{\delta_{\beta} \over (x^2 + x_{\rm c}^2)^{3\beta/2}},
\label{eqn:rho_beta}
\end{equation}
where $f_{\rm b}=\Omega_{\rm b}/\Omega_{\rm m}$, 
$x_{\rm c}=r_{\rm c}/r_{\rm vir}$ is the core radius and $\beta$
determines the asymptotic slope at large radii. We confirm in Fig.~\ref{fig:denprofgas}, 
for both our simulation models, the finding by other authors 
(e.g. Ascasibar et al. 2003, hereafter AYMG; Rasia et al. 2004, hereafter RTM)
that the model is not a good fit to the data, systematically over-predicting the 
slope of the profile at small radii and under-predicting it at large radii. 

RTM suggested an alternative expression for the density profile, which we write as
\begin{equation}
{ \rho_{\rm gas}(x) \over f_{\rm b}\rho_{\rm cr}} = 
{\delta_{\alpha} \over (x + x_{\rm p})^{\alpha} },
\label{eqn:rho_rasia}
\end{equation}
where they fixed $x_{\rm p}=0.04$ and $\alpha=2.5$ 
(shown in Fig.~\ref{fig:denprofgas} as a dotted curve).
As can be seen however, this profile does not
provide a good fit to our data. A good fit can be
achieved for our {\it Non-Radiative} clusters if we instead allow the data to
select values for $x_{\rm p}$ and $\alpha$; our results favour a steeper 
asymptotic slope  ($\alpha \sim 3$) and the ICM in
our clusters is less centrally concentrated than found by RTM. For the 
{\it Feedback} clusters, equation~\ref{eqn:rho_rasia} also improves the
fit over the $\beta$-model, but still does not provide a satisfactory description 
of the data, again getting the slope wrong at large and small radii. 
We therefore caution the use of such expressions as a general tool in
the modelling of clusters: a larger sample of clusters spanning a wider range
in mass will be required in order to investigate this issue further.



\subsection{ICM entropy profiles}

\begin{figure}
\centering
\centerline{\psfig{file=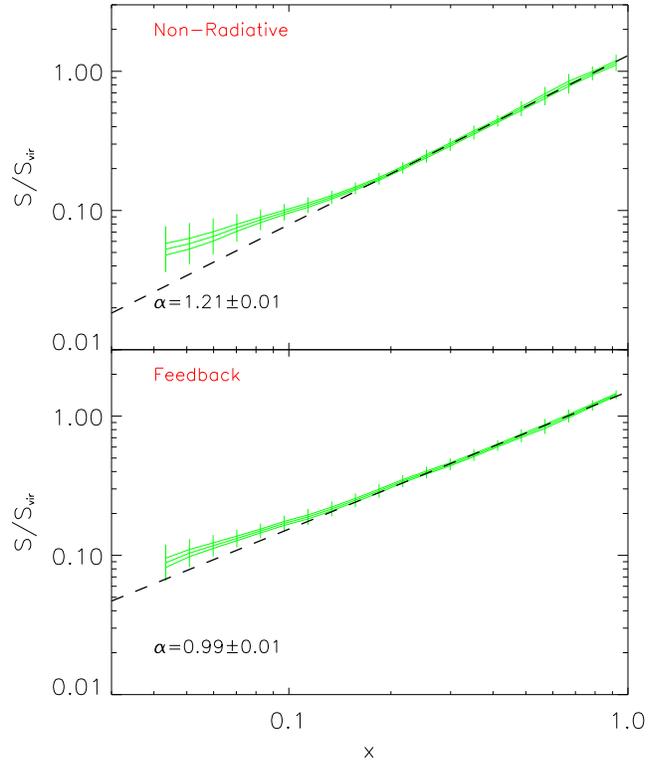,width=8.5cm}}
\caption{Dimensionless entropy profiles for clusters
in the {\it Non-Radiative} (top) and {\it Feedback} (bottom)
models. The dashed line 
is the best-fitting power law to the mean profile for $0.2<x<1$ (the
power-law index is given in the legend).}
\label{fig:entprof}
\end{figure}

\begin{figure*}
\centering
\centerline\centerline{\hbox{ \psfig{file=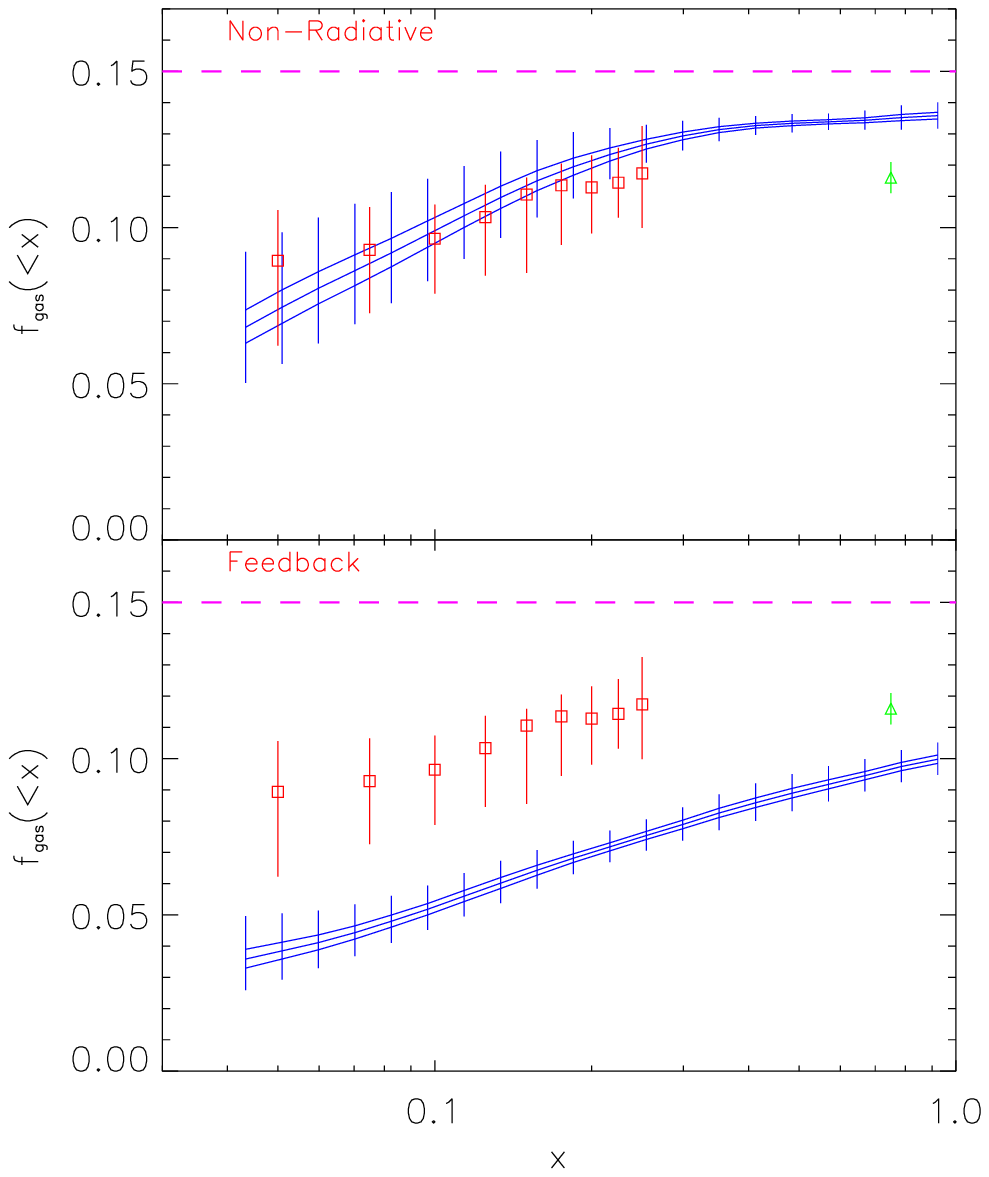,width=8.5cm}
                              \psfig{file=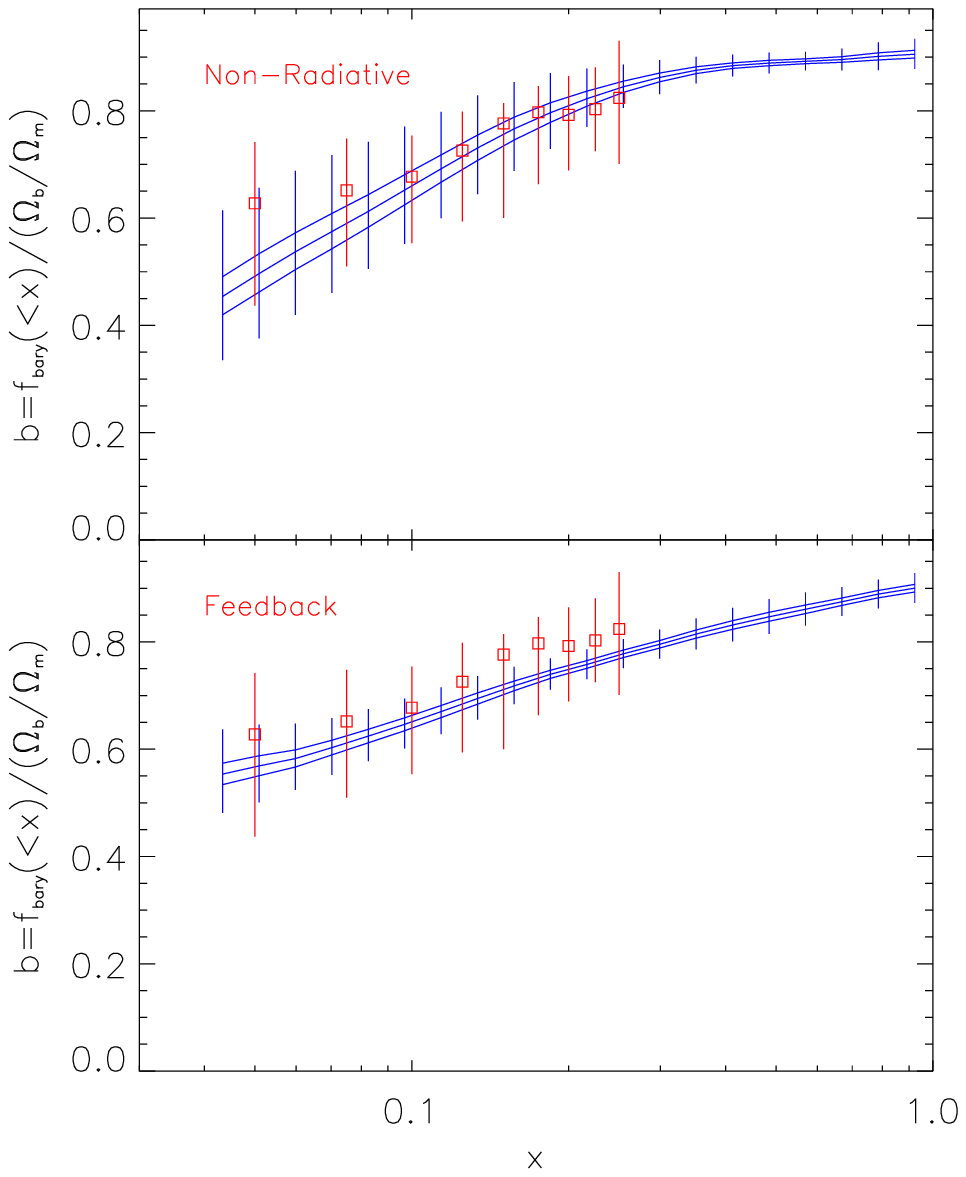,width=8.5cm} }}
\caption{ICM gas fraction (left) and baryon fraction (right) 
profiles for clusters in the {\it Non-Radiative} and
{\it Feedback} models. The global baryon fraction, 
$\Omega_{\rm b}/\Omega_{\rm m}=0.15$, 
is plotted as a horizontal dashed line. The triangle is the error-weighted
mean measurement at $r_{200}$ ($x \sim 0.75$) by Ettori (2003).
Square symbols are data points from Allen et al. (2004)} 
\label{fig:fgasprof}
\end{figure*}

The ICM entropy profile is a particularly useful quantity as it reveals
information about the nature of non-gravitational processes in clusters
(e.g. Evrard \& Henry 1991; Kaiser 1991; Bower 1997; Voit \& Bryan 2001; 
Voit et al. 2002,2003a; Ponman et al. 2003; Voit \& Ponman 2003b; Kay 2004;
Borgani et al. 2004). 
Gravitational heating models, in which the outer entropy profile 
is determined by a spherical accretion shock, predicts an outer slope of 1.1
(Tozzi \& Norman 2001). Observed entropy profiles exhibit
similar values (e.g. Arnaud, Pratt \& Pointecouteau 2004), suggesting that
the main effect of non-gravitational processes is to increase the
normalisation of the entropy profile without significantly affecting
its shape (see Ponman et al. 2003).

In Fig.~\ref{fig:entprof} we present ICM entropy profiles,
scaled by $S_{\rm vir} = kT_{\rm vir}(\mu m_{\rm H})^{2/3}(\Delta \rho_{\rm cr} 
\Omega_{\rm b} / \Omega_{\rm m})^{-2/3}$. The profiles are very close to
power laws for  $x>0.2$. Our {\it Non-Radiative} clusters
predict an outer slope of 1.2, very similar to that predicted by Tozzi \&
Norman's models, and consistent with a study of smaller systems, shown to 
be in good agreement with results from independent simulations using {\sc enzo},
an adaptive mesh refinement code (Voit, Kay \& Bryan 2004).
Cooling and feedback increase the entropy distribution of the gas 
at all radii, acting to flatten the profiles: at $x=1$ the increase is 
only 25 per cent but rises to almost a factor of 2 at $x=0.2$. Note, however,
the outer slope does not change significantly, $S \propto x$, and so is still
similar to observational determinations.

\subsection{Gas fraction profiles}
\label{subsec:gasfrac}

A unique feature of cluster-sized haloes is that, in principle, most (if not all) 
of their baryonic content can be observed directly. Combined with an estimate
of the total cluster mass, it is then possible to place constraints 
(in particular, an upper limit) on $\Omega_{\rm m}$, assuming a baryon 
density inferred from elsewhere, such as the nucleosynthetic value (White et al. 1993b). 

Panels on the left of Fig.~\ref{fig:fgasprof} illustrate 
cumulative ICM gas fraction profiles, $f_{\rm gas}(<x) = M_{\rm gas}(<x)/M_{\rm tot}(<x)$. As 
was found by previous authors (e.g. Eke et al. 1998; Frenk et al. 1999) the
{\it Non-Radiative} profile appears to converge to a value close to, but not exactly, the global
value at the virial radius (we find $f_{\rm gas} \sim 0.9 \Omega_{\rm b}/\Omega_{\rm m}$). (The
baryon fraction does not reach the global value until $x \sim 2$.)
The rise is gradual: $f_{\rm gas}(x=0.1) \sim 0.8 f_{\rm gas}(x=1)$. 

Gas fractions in the {\it Feedback} clusters are $\sim 25$ per cent lower 
at $x=1$ due to radiative cooling: as can be seen from the right half of the figure, 
the total baryon fraction profiles reach the same value as in the {\it Non-Radiative} 
clusters. The increase in gas fraction
with radius is also larger, $f_{\rm gas}(x=0.1) \sim 0.6 f_{\rm gas}(x=1)$, 
as a result of the gas being more extended due to its higher entropy.

Recently, Allen et al. (2004) measured gas fraction profiles for
26 relaxed X-ray luminous clusters observed with {\it Chandra}, out to $\Delta=2500$. 
They found that the profiles rise from $f_{\rm gas} \sim 0.07$ at $x=0$ to 
$f_{\rm gas} \sim 0.12$ at $x=0.25$ ($r_{2500}$). These results are also shown in 
Fig.~\ref{fig:fgasprof}, plotted as squares. Intriguingly, the observational data resemble 
our {\it Non-Radiative} clusters more than our {\it Feedback} clusters, with the
latter having gas fractions that are around 60 per cent of the observed value at 
$r_{2500}$.
We also show the error-weighted mean gas fraction at $r_{200}$ ($x \sim 0.75$), 
as measured by Ettori (2003), consistent with the result of Allen et al. if the
profiles do not continue to rise much beyond $r_{2500}$. Here, our {\it Feedback}
results are approximately 80 per cent of the observed estimate.

In the right-hand panels of Fig.~\ref{fig:fgasprof} we plot {\it bias} factor profiles, 
$b = f_{\rm bary}/(\Omega_{\rm b}/\Omega_{\rm m})$, i.e. the baryon fraction, $f_{\rm bary}$,
in units of the global value. Allen et al. assume a constant fraction (16 per cent)
of baryons in stars. Interestingly, the {\it Feedback} clusters are now in good agreement
with their observations. Cooling leads to a rise in the central baryon fraction, 
flattening the profile with respective to the {\it Non-Radiative} case. Clearly
however, our {\it Feedback} clusters have too much material in stars at all
radii ($f_{*}$ decreases monotonically with radius, from 0.54 at $x=0.05$ to 0.25 at
$x=1$). At least part of the problem may be due to forming stars at artificially low 
densities: while increasing the star formation density threshold presents other problems 
(in particular, the propagation of feedback energy to lower density regions),
it will allow more gas to be ablated as it moves through the ICM, therefore reducing
the amount of material available to form stars.

\section{Temperature profiles}
\label{sec:temprof}

A more accurate description of the ICM requires the spatial distribution
of its temperature to be measured, which can be achieved using X-ray 
spectroscopy. Both observations and simulations concur that the ICM is 
non-isothermal, and so measuring temperature gradients improves the 
accuracy of estimating a cluster's mass if it is in hydrostatic equilibrium.

Early observations of azimuthally-averaged temperature profiles
debated whether the inner profile was declining 
with radius (Markevitch et al. 1998a) or was isothermal (Irwin, Bregman \& Evrard 1999;
White 2000; Irwin \& Bregman 2000). A later study by De Grandi \& Molendi 
(2002) was consistent with isothermality outside cooling-flow regions
and within $\sim 0.2 r_{\rm vir}$. This was confirmed
by Allen et al. (2001) for their sample of 6 luminous relaxed clusters
observed with {\it Chandra}, and more recently by Arnaud et al. (2004) for 7 clusters
observed with {\it XMM-Newton}. Although the
situation is still not robust, the general consensus is that the ICM 
temperature rises from the centre to $\sim 0.1r_{\rm vir}$, is approximately isothermal
then starts to decline beyond $\sim 0.2-0.3r_{\rm vir}$. More high-quality data from 
{\it XMM-Newton} and {\it Chandra} will improve the situation considerably.

\subsection{3D temperature profiles}

\begin{figure}
\centering
\centerline{\psfig{file=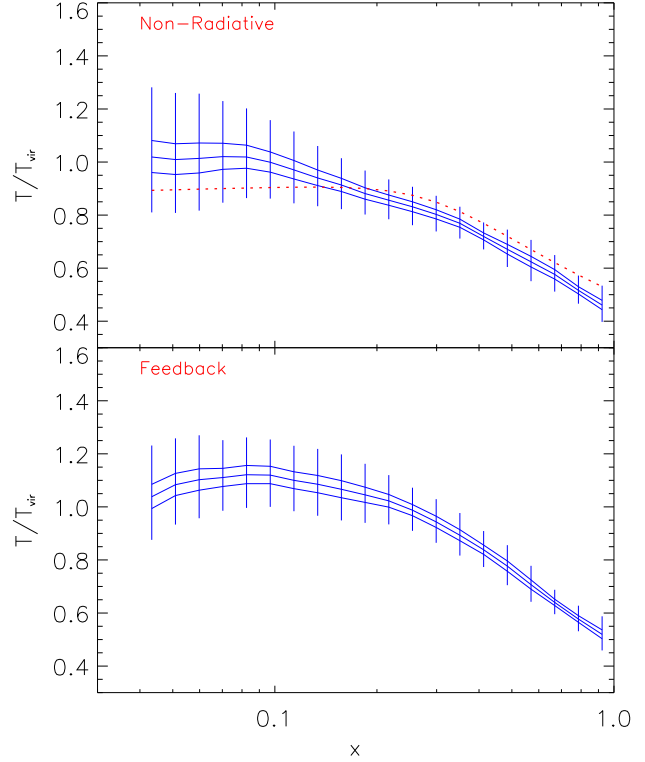,width=8.5cm}}
\caption{Dimensionless temperature profiles for clusters
in the {\it Non-Radiative} and {\it Feedback} models. 
The best-fitting 
profile to the results of Rasia et al. (2004) is shown as a dotted curve.}
\label{fig:temprof}
\end{figure}

We first present 3D temperature profiles, shown in Fig.~\ref{fig:temprof}.
Each profile is scaled by the isothermal model virial temperature of the cluster, 
$T_{\rm vir}=G \mu m_{\rm H} M_{\rm vir}/2 k R_{\rm vir}$. 
For the {\it Non-Radiative} clusters, our results are in good agreement
with previous numerical studies (e.g. Evrard 1990; Navarro et al. 1995;
Eke et al. 1998), i.e. the profile slowly varies out to $x \sim 0.2$,
then declines by around a factor of 2 out to $x=1$. A more contentious
issue is whether the inner profile is isothermal (e.g. RTM)
or continues to rise to the centre (e.g. AYMG). We
find our average {\it Non-Radiative} profile is isothermal within 
$x \sim 0.1$, although the core temperature is hotter than that
found by RTM (their best-fit profile is shown as a dotted curve). 
Given AYMG also used the entropy-conserving version
of {\sc gadget} (RTM did not), the difference in the slope 
of the inner profile could be due to the selection of objects (AYMG studied 
lower-mass systems), although we note both RTM and AYMG performed studies
with both higher mass and force resolution than ours.

Including cooling and feedback produces a modest increase ($\sim 10$ per cent
at $x=1$) in the temperature of the ICM. Note also that the average {\it Feedback} 
profile begins to turn over within $x \sim 0.1$ due to radiative cooling.


\subsection{Projected temperature profiles}

\begin{figure*}
\centering
\centerline{\psfig{file=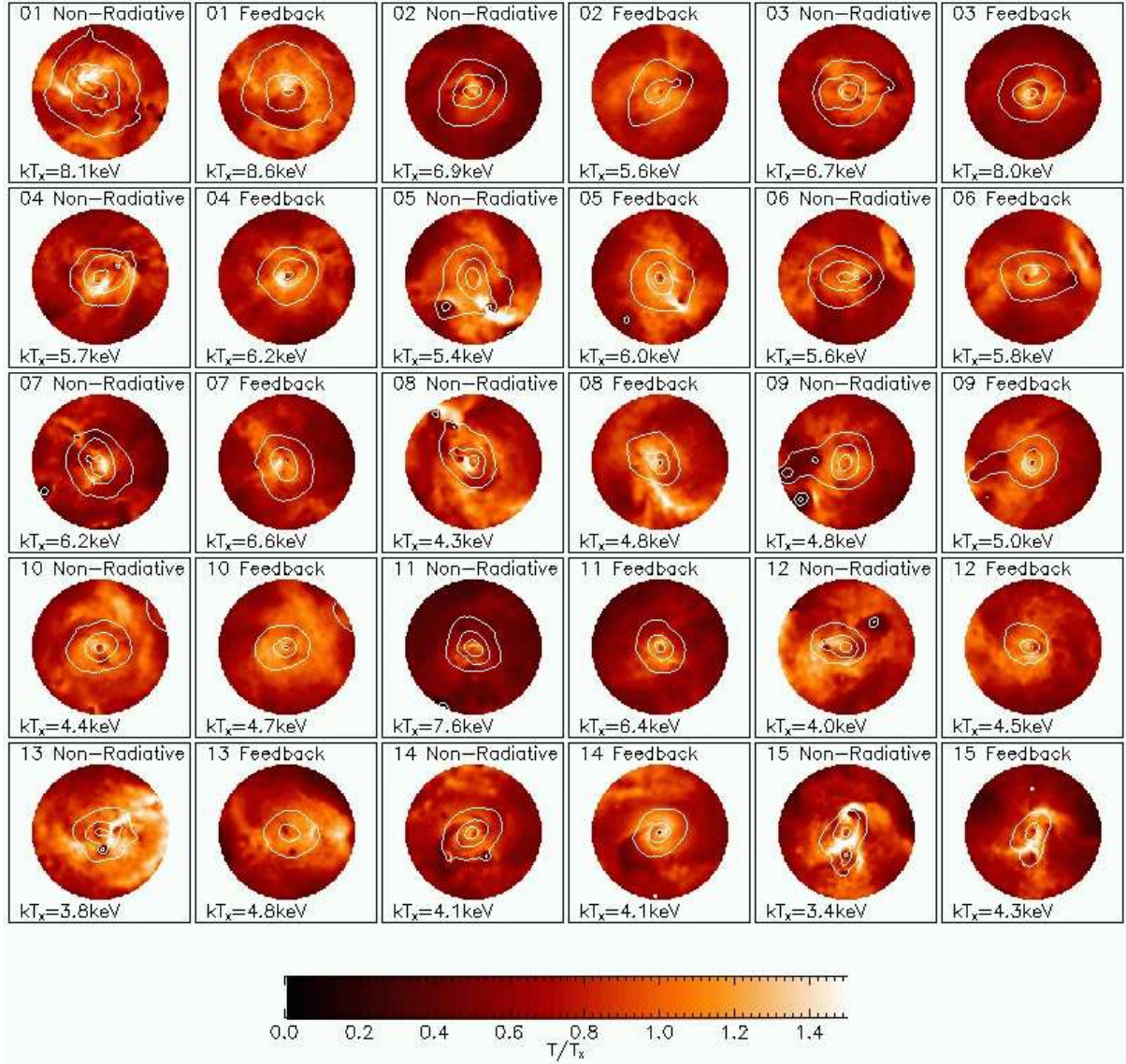,width=16cm}}
\caption{X-ray temperature maps for the clusters studied in
this paper. Maps are shown within an Abell Radius ($1.5 \hMpc$). 
X-ray surface brightness contours (with order of magnitude
intervals) are overlaid.}
\label{fig:tmap}
\end{figure*}

To compare our temperature profiles with observations, we have
constructed projected profiles. 
Here, we emulate the method employed by Loken et al. (2002), 
who themselves tried to mimic the procedure employed by Markevitch 
et al. (1998a). X-ray emission-weighted temperature and surface brightness
maps were first constructed for the $1.5-11$keV 
band, using the procedure 
detailed by Onuora, Kay \& Thomas (2003). Each map was constructed from
a cube of width $4 \hMpc$, centred on the cluster, and the length of the 
pixels were set to the gravitational softening, $\epsilon=0.02\hMpc$
(hence each map is $200 \times 200$ pixels). Maps were then
centred on the pixel containing the maximum surface brightness and normalised
by the X-ray emission-weighted temperature ($T_{\rm X}$) within $1\hMpc$ of this centre.

The temperature maps (with surface brightness contours overlaid) are shown in
Fig.~\ref{fig:tmap}, out to one Abell radius (1.5$\hMpc$). It is clear from this
figure (see also Loken et al. 2002; Onuora et al. 2003) that the temperature of
the ICM is very asymmetric and relates to the individual merger history of each
cluster. How much of this information can be recovered from observed clusters, however,
is a question that is now only starting to be addressed (e.g. Gardini et al. 2004;
Mazzotta et al. 2004).

\begin{figure*}
\centering
\centerline{ \hbox{ \psfig{file=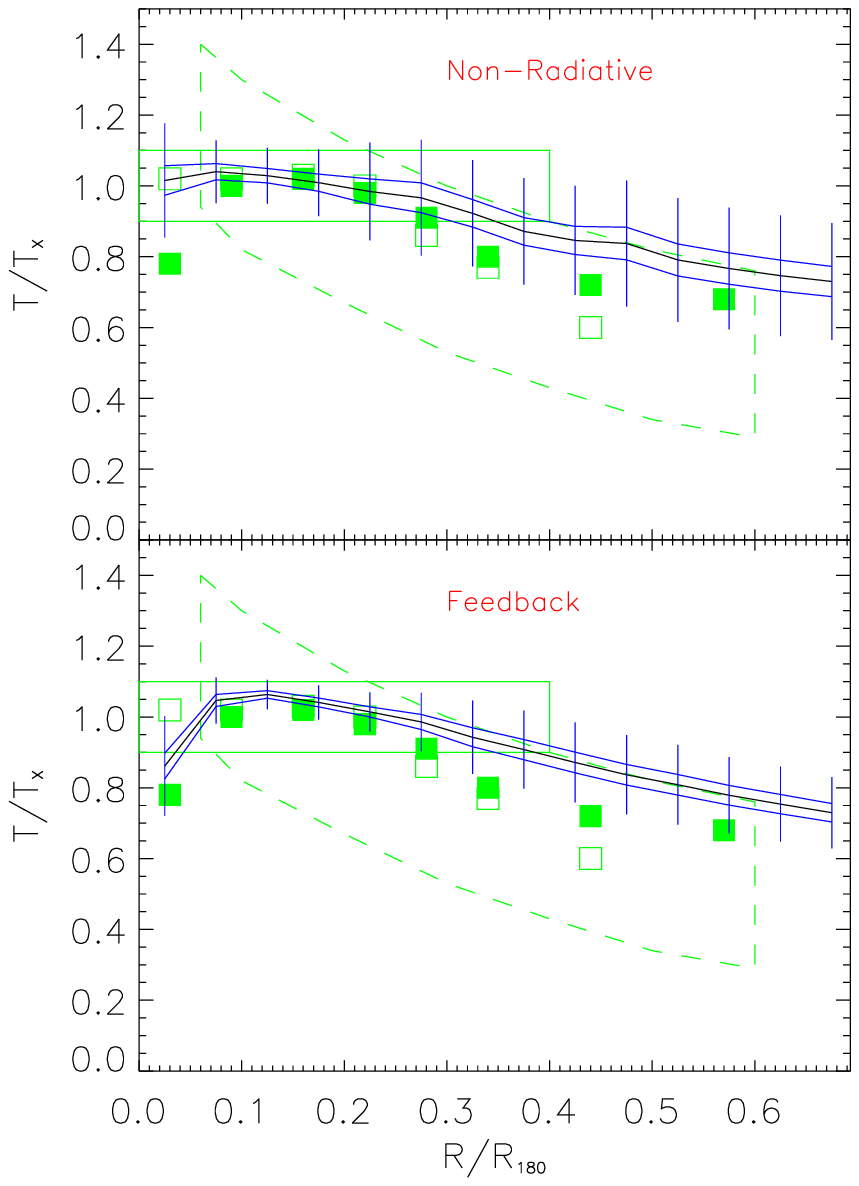,width=8.5cm}
                    \psfig{file=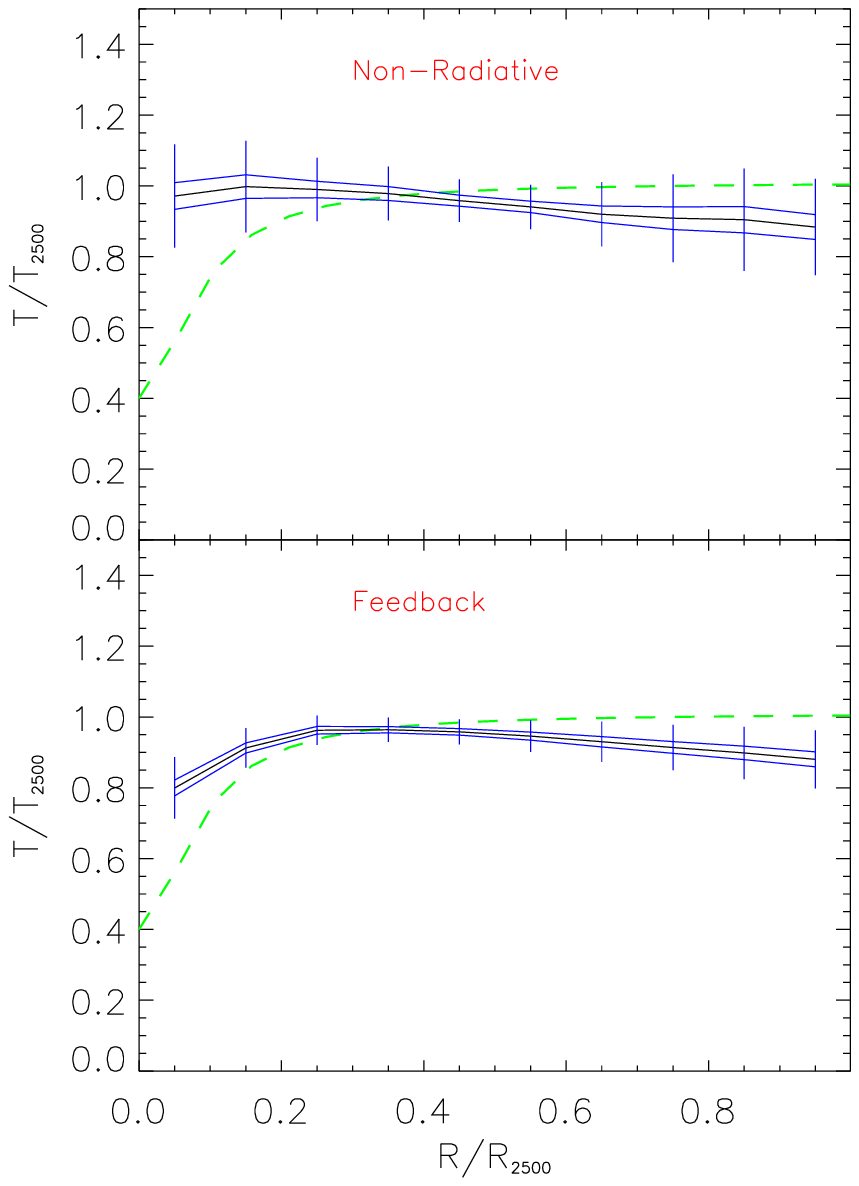,width=8.5cm} }}
\caption{Projected ICM temperature profiles compared with various
observational data. Left panels: profiles are plotted out to $0.7R_{180}$
and compared with results from Markevitch et al. (1998a; dashed box region),
De Grandi \& Molendi (2002; open and filled squares) and
Arnaud, Pratt \& Pointecouteau (2004; solid rectangle).
Right panels: profiles are plotted out to $R_{2500}$ and compared to
the result from Allen et al. (2001).}
\label{fig:ptprof}
\end{figure*}

We then constructed projected emission-weighted temperature profiles by 
azimuthally averaging temperatures within bins of width 0.05$R_{180}$, where
$R_{180}=1.95(kT_{\rm X}/10{\rm keV})^{1/2}\hMpc$ is the radius commonly
used by observers, calibrated from the simulations of Evrard et al. (1996).
The results are shown in the top panels of 
Fig.~\ref{fig:ptprof}, compared to the data of Markevitch et al.
(1998a); De Grandi \& Molendi (2002) and Arnaud et al. (2004). 
Intriguingly, our mean profiles are in reasonably good agreement with
the observations within $0.2R_{180}$ (excluding the cooling-flow
region in the {\it Non-Radiative} model) although they fall off less rapidly
in the outer parts. This is contrary to recent claims 
by several groups (e.g. Loken et al. 2002; AYMG;
Borgani et al. 2004), who found consistency with the observations at large radii 
but whose profiles continued to rise into the centre. For example,
Borgani et al., who included a model for both cooling and feedback in their 
calculation, found that the average temperature of clusters above 3keV
increased to 1.2 times the emission-weighted temperature at $\sim 0.3R_{180}$, before
turning over. It is unclear whether the discrepancy is due to numerical 
resolution effects (Borgani et al. simulated their clusters at higher resolution)
or differences in the feedback model or both. It is imperative that such discrepancies
are fully understood and we aim to do so in future work.

In the bottom panels of Fig.~\ref{fig:ptprof}, we also 
compare projected mass-weighted profiles to the result
found by Allen et al. (2001), plotted within $R_{2500}$. 
Our {\it Feedback} model is in reasonably good agreement at
these radii (to within $\sim 10$ per cent).

\section{Mass Estimates}
\label{sec:mest}

\begin{figure}
\centering
\centerline{\psfig{file=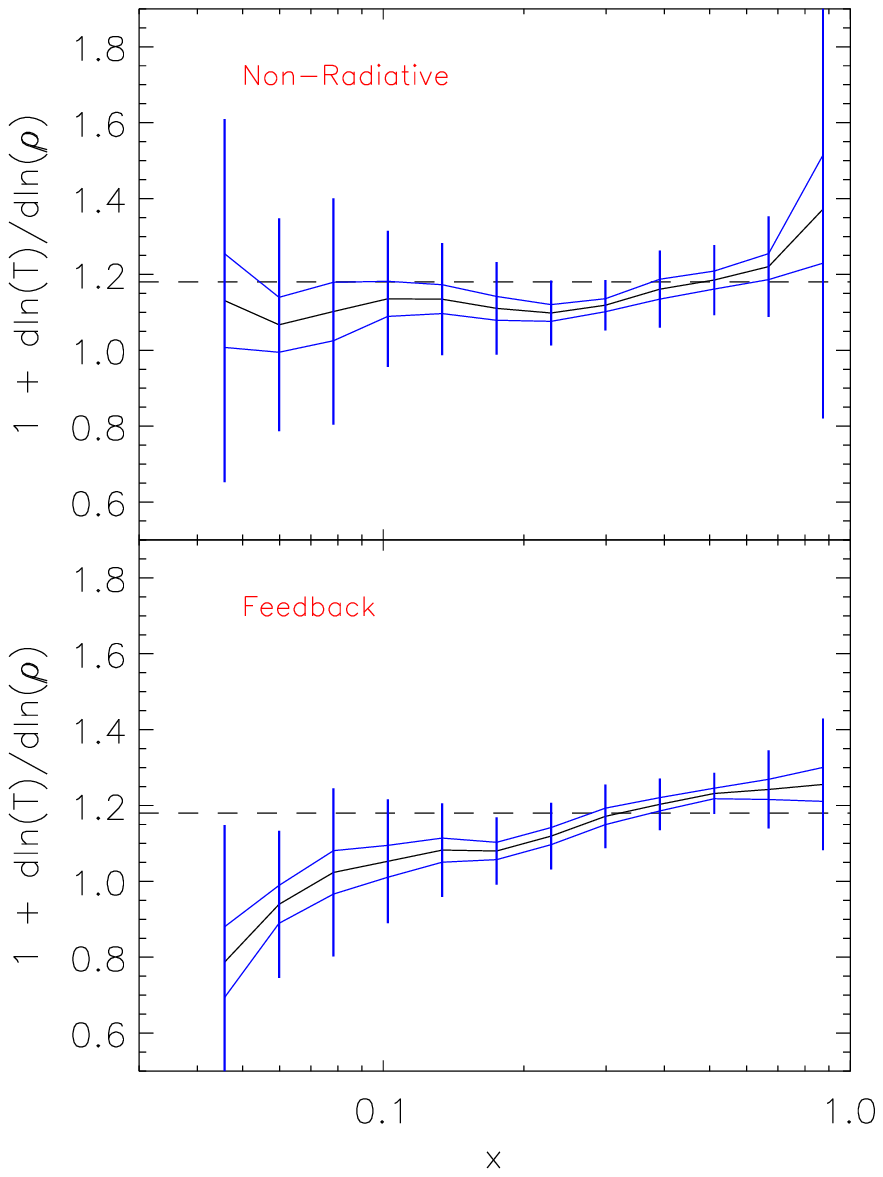,width=8.5cm}}
\caption{Effective polytropic index, $\gamma$, as a function of radius
for the {\it Non-Radiative} and {\it Feedback} models. The horizontal dashed
line marks the value $\gamma=1.18$, found to be a good description for
the relaxed poor clusters studied by Ascasibar et al. (2003).} 
\label{fig:polyindex}
\end{figure}

We now investigate the accuracy with which our cluster masses can be
recovered when making various assumptions regarding the radial 
structure of the ICM. If the ICM is in hydrostatic equilibrium, the
mass can be calculated from
\begin{equation}
M(<x) = - { (x \, r_{\rm vir})^2 \over G \rho(x) } \, {{\rm d}P \over {\rm d}r}.
\label{eqn:hydrostat}
\end{equation}
The simplest model would then be to assume the ICM is a polytropic gas,
$P=\kappa\rho^{\gamma}$. For isothermal distributions, $\gamma=1$, 
and for adiabatic distributions, $\gamma=5/3$. We show in 
Fig.~\ref{fig:polyindex} (plotting the effective polytropic index, 
$\gamma=1+d\ln(T)/d\ln(\rho)$, at each radius), that this assumption is 
not valid in general for our simulated clusters. For the {\it Non-Radiative} systems, 
$\gamma \sim 1.1$ out to $x \sim 0.2$ then increases to $\gamma \sim 1.3$
at $x=1$. For the {\it Feedback} systems, $\gamma$ increases monotonically
from the centre outwards, varying from $\sim 0.8$ at $x=0.04$ to $\sim 1.3$ at $x=1$. 
Note in this case, $\gamma$ is less than unity in the cluster core,
reflecting the positive radial temperature gradient induced by cooling.

AYMG split their sample of poor clusters into relaxed, minor and major merger systems and
found the first two to be well described by a single polytrope
with $\gamma=1.18$ (shown in our figure as a horizontal dashed line). 
For their major mergers, the main deviation in $\gamma$ was in the central
regions ($x \approxlt 0.3$), where the gas was close to isothermal. 
On average, our {\it Non-Radiative} clusters also have lower values of
$\gamma$ in their centres, suggesting that our objects are less relaxed than theirs, 
as would be expected given their higher mass and hence longer dynamical times.

\subsection{The isothermal $\beta$-model}

\begin{figure}
\centering
\centerline{\psfig{file=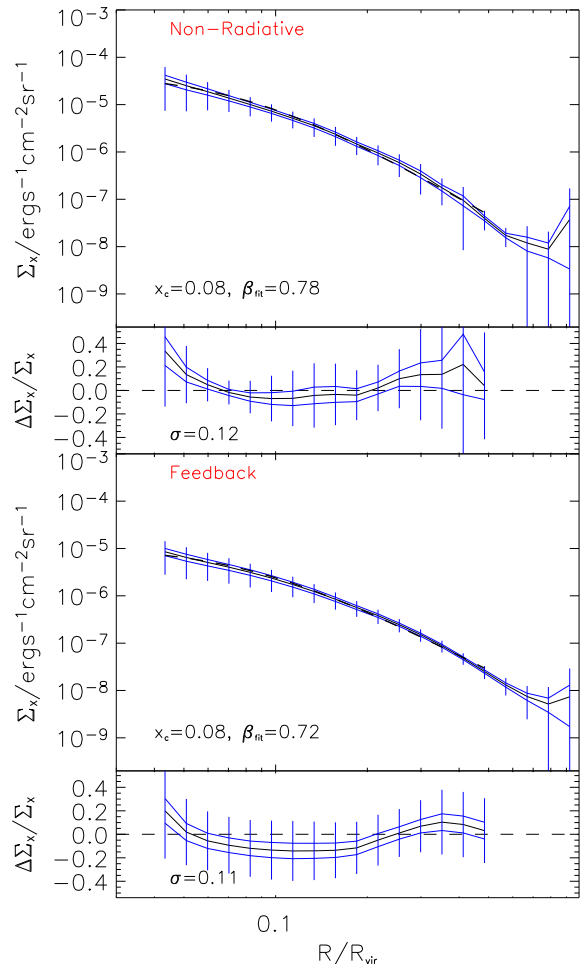,width=8.5cm}}
\caption{X-ray surface brightness profiles for clusters in
the {\it Non-Radiative} and {\it Feedback} models. Fits
are only produced for $R<0.5R_{\rm vir}$.}
\label{fig:sbprof}
\end{figure}

Although our clusters are clearly not isothermal, it is nevertheless
interesting to apply the isothermal $\beta$-model (Section~\ref{subsec:denprof})
to equation~\ref{eqn:hydrostat}
as it is this model that is most commonly used by observers to estimate
cluster masses. First, the X-ray surface brightness profile is fitted
\begin{equation}
\Sigma_{\rm X}(R) = {\Sigma_{\rm X}(0) \over (R^2 + R_c^2)^{3\beta_{\rm fit}-1/2}},
\label{eqn:sb_beta}
\end{equation}
where $R_c$ is the core radius and $\beta_{\rm fit}$ is the asymptotic slope
of the profile at large radii. The results from applying the same procedure 
to our surface brightness maps are shown in Fig.~\ref{fig:sbprof}. As has
already been shown when fitting the ICM density profile (Fig.~\ref{fig:denprofgas}),
it is clear that this model does not describe the simulation data well over the 
whole range of radii. We only fit the data for $x<0.5$ (at larger radii
the profiles diverge due to infalling substructure); even then, the average
fractional difference is around 10 per cent. 

\begin{figure}
\centering
\centerline{\psfig{file=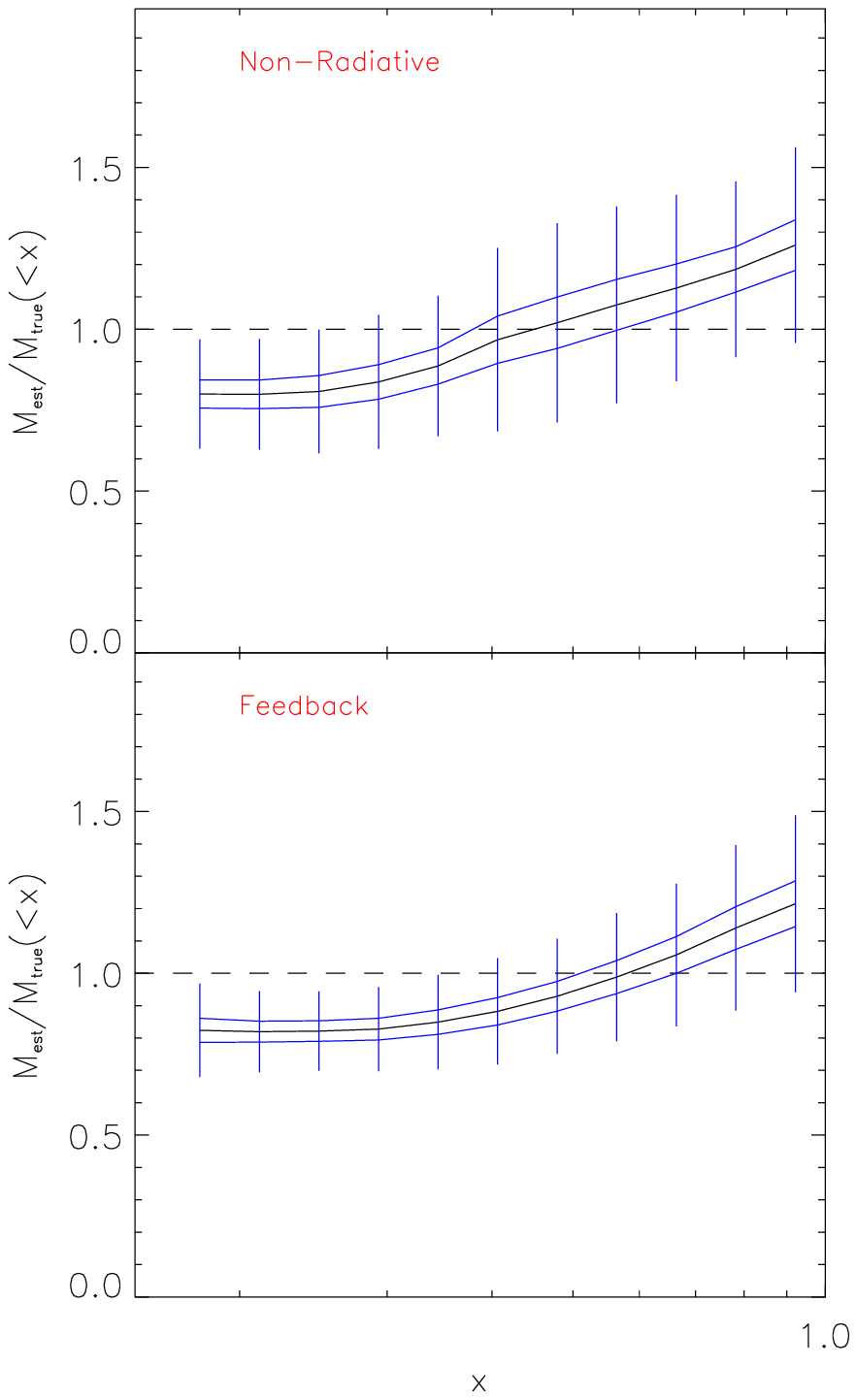,width=8.5cm}}
\caption{Ratio of estimated to true total cluster mass 
within a given radius using the isothermal $\beta$-model.}
\label{fig:beta_mass}
\end{figure}

The total mass within a given radius can then be estimated
\begin{equation}
M(<R) = \beta_{\rm fit} \, \left({3kT_{\rm X} \over G \mu m_{\rm H}}\right) \, {R^3 \over R^2 + R_c^2},
\label{eqn:mass_beta}
\end{equation}
where $T_{\rm X}$ is a single (emission-weighted) measurement of the ICM temperature.
In Fig.~\ref{fig:beta_mass} we show results from applying equation~\ref{eqn:mass_beta}
to our clusters, plotting the ratio of the estimated to the true mass within a given
radius. The results are very similar for the
{\it Non-Radiative} and {\it Feedback} clusters: equation~\ref{eqn:mass_beta}
underestimates the mass by $\sim 20$ per cent at $x<0.3$ and overestimates the
mass by $\sim 30$ per cent at $x=1$. Interestingly, at $x=0.5$
($\sim R_{500}$) the estimated mass is within 5 per cent of the true mass.
Within $R_{200}$ ($x \sim 0.75$), 
masses can be estimated to better than 20 per cent.
A similar study was performed by Evrard et al. (1996), who considered
various models and found a similar trend with radius, albeit better agreement between
the estimated and true masses at $x<0.5$.

\subsection{Single versus spatial temperature models}

In the new era of {\it Chandra} and {\it XMM-Newton}, obtaining spatially resolved
density and temperature information is now becoming possible. It is therefore of 
interest to estimate cluster masses when accurate descriptions for their
density and temperature profiles are known.

\begin{figure*}
\centering
\centerline{\psfig{file=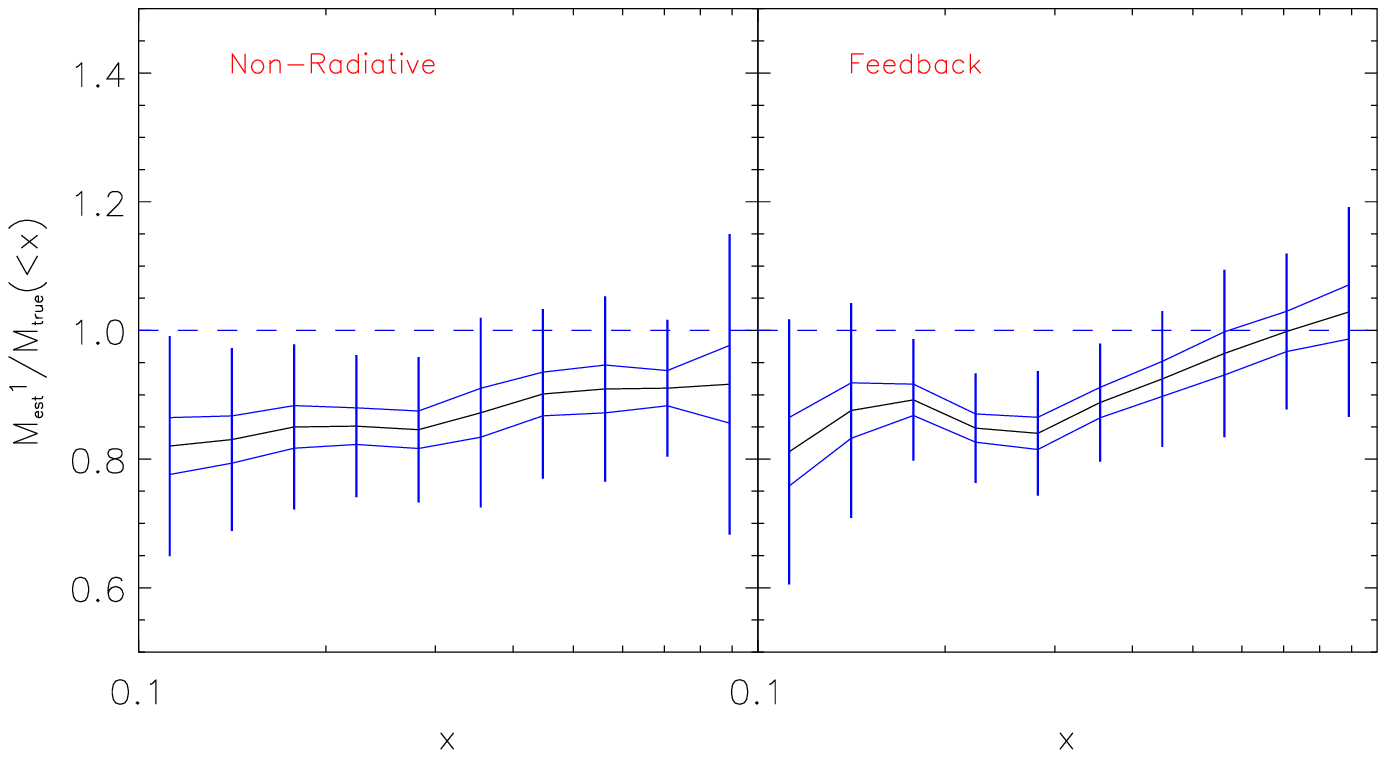,width=12cm}}
\centerline{\psfig{file=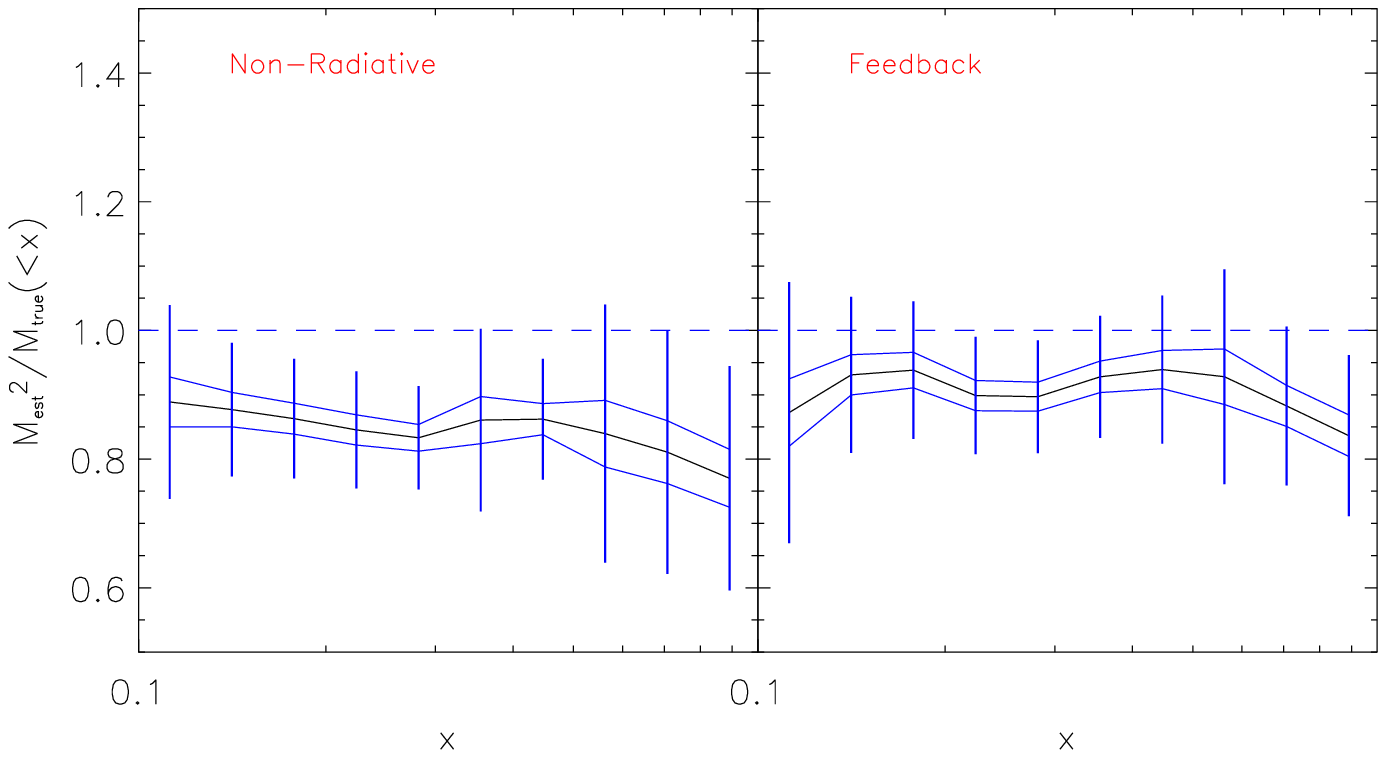,width=12cm}}
\centerline{\psfig{file=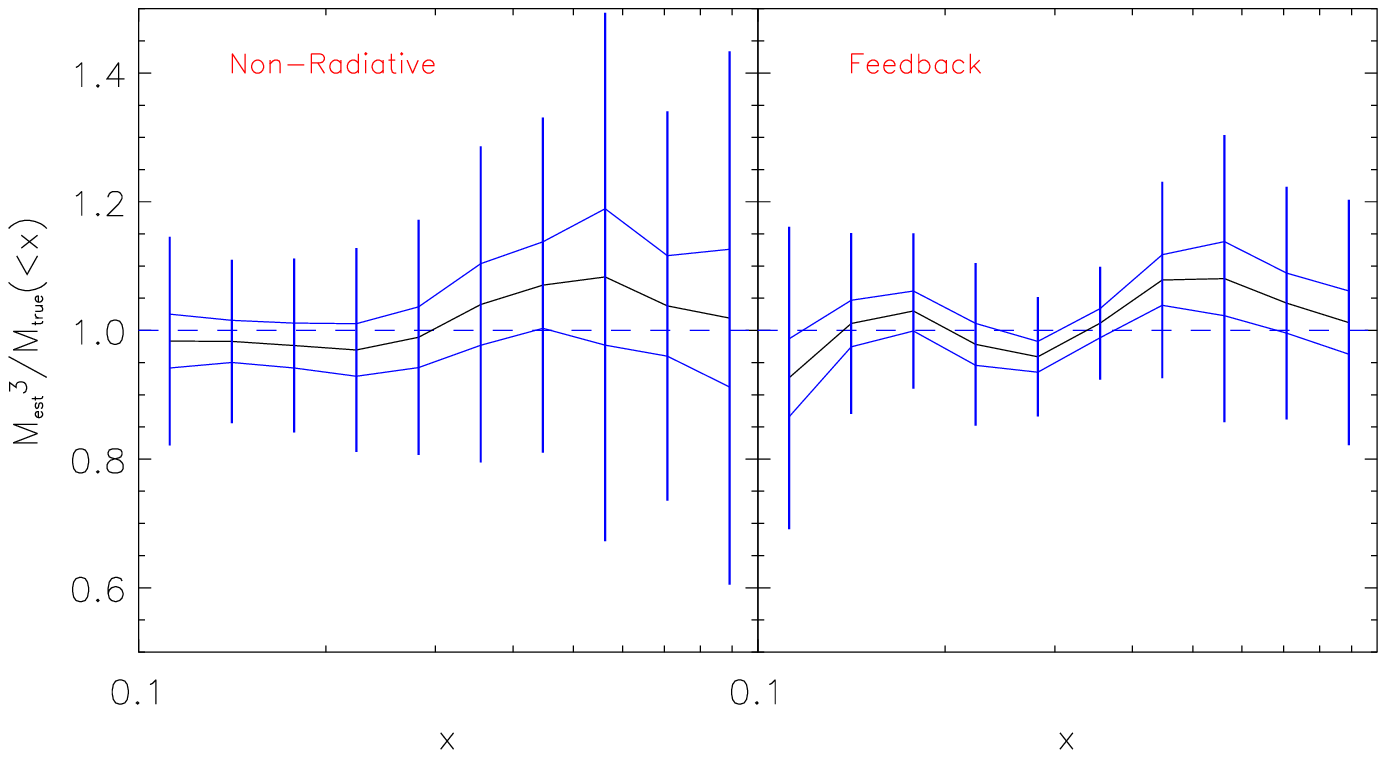,width=12cm}}
\caption{Ratios of estimated to true cluster mass as a function
of radius, excluding
(top panels) and including (middle panels) spatial temperature
information. Results in the bottom panels are when additional,
isotropic velocity dispersion information is also included.}
\label{fig:mrat}
\end{figure*}

Following RTM, we first perform the test where we use a single temperature
to describe the ICM but use its true density profile. 
Equation~\ref{eqn:hydrostat} becomes
\begin{equation}
M_{\rm est}^{1} (<x) =  {k T(<x) \, x \, r_{\rm vir} \over G \mu m_{\rm H} }
                         \, { d\ln\rho(x) \over d\ln(x)} ,
\end{equation}
where $T(<x)$ is the mass-weighted temperature within $x$.
Results are shown in Fig.~\ref{fig:mrat}, again plotting the estimated
to true mass ratio as a function of radius. 

Our {\it Non-Radiative} result is in reasonably good agreement with the findings of RTM, 
with the estimated mass being lower, but within 20 per cent of the true mass for
$0.1 < x < 1$. At $x<0.3$, ratios for the {\it Feedback} clusters are similar,
but the estimated mass approaches the true mass much more rapidly at larger radius, 
being within $\sim 5$ per cent between $x=0.5$ ($R_{500}$) and $x=1$. 
The integrated mass-weighted temperature
is $\sim 10$ per cent larger in the {\it Feedback} clusters at all radii, increasing
the estimated mass by the same factor. This is cancelled out at small
radii by the flatter inner slope of the density profile. At larger radii however,
the slope increases with radius, eventually approximating the same slope as
in the {\it Non-Radiative} clusters and causing the mass estimate to rise with
radius.

We now include spatial temperature information in our mass estimate
\begin{equation}
M_{\rm est}^{2} (<x) =  {k T(x) \, x \, r_{\rm vir} \over G \mu m_{\rm H} }
                         \, \left[ {d\ln\rho \over d\ln x} + {d\ln T \over d\ln x} \right],
\end{equation}
with results shown in the middle panels of Fig.~\ref{fig:mrat}.
Our {\it Non-Radiative} results are again in good agreement with RTM.
Both models underpredict the mass at all radii by up to 20 per cent. Again,
the mass estimate is slightly higher for the {\it Feedback} clusters because 
of the increase in ICM temperature. Clearly from these results, our clusters
cannot be completely in hydrostatic equilibrium. As pointed out by RTM, the isothermal
mass estimator is more accurate than the non-isothermal case at larger
radii, because the integrated temperature happens to cancel out
the effects of gas motion.

\begin{figure}
\centering
\centerline{\psfig{file=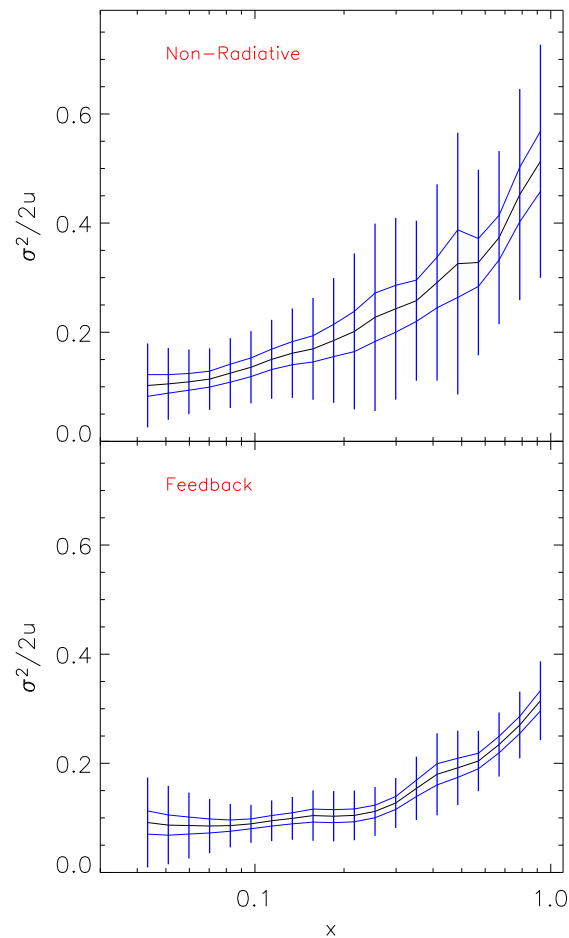,width=8.5cm}}
\caption{Ratio of kinetic to thermal energy, $\sigma^2/2u$, for the 
{\it Non-Radiative} and {\it Feedback} clusters.}
\label{fig:vrad}
\end{figure}

The lack of agreement between estimated and true cluster masses is due
to the presence of kinetic support, which becomes particularly significant
at large radii. Radial velocity profiles confirm that our clusters are,
to very good approximation, virialised systems. However, as
Fig.~\ref{fig:vrad} shows, the velocity dispersion of the ICM is 
non-negligible and becomes increasingly significant with radius. 
Note, however that kinetic support is less important for the 
{\it Feedback} clusters: the {\it Non-Radiative} clusters
have a positive velocity dispersion gradient at all radii, whereas in the 
{\it Feedback} clusters, a significant gradient appears at $x>0.2$. 
Thus, the ICM in the {\it Feedback} clusters is hydrostatic to better approximation
than in the {\it Non-Radiative} clusters.

Including an isotropic velocity dispersion term in equation~\ref{eqn:hydrostat},
namely $P = \rho (kT/\mu m_{\rm H} + \sigma^2/3)$, we show ratios of estimated
to true mass estimates in the bottom panels of Fig.~\ref{fig:mrat}. On average,
the estimated masses are now in good agreement with the true masses.

\section{Conclusions}
\label{sec:conclusions}

In this paper we have studied high-resolution $N$-body/hydrodynamical
simulations of galaxy clusters to investigate the effects of
radiative cooling and galactic feedback on the properties of the
ICM. In particular, we compared a set of {\it Non-Radiative} simulations
(the simplest model that is well-studied in the literature, but is nevertheless
ruled out by X-ray observations of clusters) to our {\it Feedback} model
where a fraction (10 per cent) of cooled material is given a fixed amount
of entropy ($1000 \kevcm2$) that is then redistributed in the ICM. 
The latter model reproduces the observed cluster $L_{\rm X}-T_{\rm X}$ relation,
signifying the correct level of excess entropy in the ICM core. Our main
conclusions are as follows.

\begin{itemize}

\item The distribution of total mass is, on average, reasonably well 
described by an NFW profile, with residual differences within 5 per cent.
Including cooling and feedback does not significantly alter the shape of the 
cluster potential, causing only a $\sim 10$ per cent increase in the best-fit
concentration parameter.

\item Cooling and feedback act to increase the entropy of the ICM
everywhere, although the effect is more pronounced within the central
region, where the cooling rate (and hence the rate of mass/energy
injection) is higher. Outside $\sim 0.2 r_{\rm vir}$, the entropy 
profile scales approximately linearly with radius, consistent with
recent X-ray observations.

\item As a consequence of the excess entropy, the ICM density profile
is lower and flatter in the {\it Feedback} clusters. 
The isothermal $\beta$-model does not provide a good fit to the
ICM density profile in either numerical models, systematically overpredicting
the inner slope and underpredicting the outer slope. 

\item Baryon fractions rise gradually from the centre outwards, reaching
$\sim 90$ per cent of the global value ($\Omega_{\rm b}/\Omega_{\rm m}$)
at the virial radius. Virtually no gas is lost from the {\it Feedback}
clusters due to galactic outflows; the lower ICM gas fraction can be accounted
for by the fraction of stars. However, the 
{\it Feedback} model does not contain enough gas within $r_{2500}$ to
agree with the observed determinations by Allen et al. (2004), although
the overall baryon fraction profiles are in good agreement. Our model
overproduces the mass in stars at all radii, possibly a symptom of
having such a low density threshold for star formation.

\item Cooling and feedback produce only a modest increase in the temperature
of the ICM (around 10 per cent at the virial radius) and cooling induces
a positive radial temperature gradient for $r<0.1r_{vir}$. The temperature
of the ICM is then primarily driven by gravity. Projected temperature
profiles are in reasonable agreement with the observations although the
decline at larger radii is not as severe.

\item On average, the ICM cannot be described by a single polytropic
equation of state: the effective polytropic index, 
$\gamma = 1 + d\ln(T)/d\ln(\rho)$ increases with radius, reaching
$\gamma \sim 1.3$ at the virial radius. Central values are 
lower than unity in the {\it Feedback} clusters, due to the positive radial 
temperature gradient induced by radiative cooling.

\item Estimating the total mass of clusters by fitting surface
brightness profiles with the isothermal
$\beta$-model is accurate to within 20-30 per cent (depending on the 
outer radius chosen), regardless of the simulated model. The model
is most accurate around $r=0.5r_{\rm vir}$ ($\sim r_{500}$), where the
error is $\sim$ 5 per cent.

\item As found by Rasia et al. (2004), the ICM is not completely hydrostatic, 
leading to an underestimate in the true mass of the cluster of up to 20 per cent 
in our {\it Non-Radiative} clusters, even when full density and temperature information 
is included in the calculation. Including a kinetic pressure term recovers the true
mass. The degree of thermalisation in the ICM improves when cooling and feedback 
are included, predicting a hydrostatic mass that is within 10 per cent of the true mass for 
$r<r_{500}$.

\end{itemize}

In summary, we find that cooling and feedback mainly affect the
inner structure of the ICM, as a consequence of the excess entropy
produced by these processes. Although our {\it Feedback} model
approximately achieves the overall desired entropy excess in clusters, it
does not provide a detailed match to the inner structure of the ICM,
namely the gas mass fraction within $r_{2500}$ is around 60 per cent
of the observed value (Allen et al. 2004).
Whether this discrepancy can be rectified within the context of our
feedback model (by varying the model parameters and/or increasing the
numerical resolution of the simulations), or whether our simulations are 
still missing some fundamental physical process, requires further investigation.

\section*{Acknowledgments}
We thank the referee, Stefano Borgani, for his insightful
comments that improved the quality of the manuscript.
We also thank Volker Springel for generously allowing us to
use {\sc gadget2} before its public release and for
his help with various technical issues, and
Steve Allen for providing observational data. Simulation data were 
generated using COSMA, the 670-processor COSmology MAchine 
at the Institute for Computational Cosmology in Durham. The 
work presented in this paper was carried out as part of the 
programme of the Virgo Supercomputing Consortium 
({\tt http://www.virgo.dur.ac.uk}). 
STK and ARJ are supported by PPARC, FRP is a PPARC Advanced Fellow.

\label{lastpage}

\end{document}